\documentclass[12pt]{iopart}

\usepackage{graphicx}
\usepackage{amssymb}

\begin{document}

\title{Approximate semi-analytical solutions for the steady-state expansion of a contactor plasma}

\author{E. Camporeale$^1$, E.~A. Hogan$^2$, E.~A. MacDonald$^3$}
\address{1 Center for Mathematics and Computer Science (CWI), Amsterdam, Netherlands.}
\address{2 University of Colorado, Boulder, CO, USA }
\address{3 NASA Goddard Space Flight Center, Greenbelt, MD, United States }
\ead{e.camporeale@cwi.nl}

\begin{abstract}
We study the steady-state expansion of a collisionless, electrostatic, quasi-neutral plasma plume
into vacuum, with a fluid model. We analyze approximate semi-analytical
solutions, that can be used in lieu of much more expensive numerical solutions. 
In particular, we focus on the earlier studies presented in \textit{Parks and Katz (1979)} \cite{parks79}, \textit{Korsun and Tverdokhlebova (1997)} \cite{korsun97}, and \textit{Ashkenazy and Fruchtman (2001)}  \cite{ashkenazy01}. 
By calculating the error with respect to the numerical solution, we can judge the range of validity for each solution.
Moreover, we introduce a generalization of earlier models that has a wider range of applicability, in terms of plasma injection profiles.
We conclude by showing a straightforward way to extend the discussed solutions to the case of a plasma plume injected with non-null azimuthal velocity.
\end{abstract}
\jl{25}
\submitted
\maketitle
\section{Introduction}\label{intro}
The modeling of the expansion of a plasma plume in the vicinity of a spacecraft has been intensively investigated in the last several years. 
In particular, in the context of space electric propulsion systems, such as ion or Hall thrusters,
a correct characterization of the emitted plasma plume is of crucial importance to avoid interactions between the energetic particles and spacecraft surface,
and thus to potentially prevent severe damages to the spacecraft \cite{boyd04,zhong08,yan12,dannenmayer13}.
Similarly, in plasma contactor technology (for instance, electrodynamics tether applications), it is important to
predict the shape and geometry of the plasma plume emitted by a spacecraft \cite{wilbur88,sanmartin10}.\\
Due to the large difference in density between the plasma plume and the background magnetospheric plasma, the physics of the plume expansion
is often well described by the expansion of plasma in vacuum, which has also been investigated thoroughly for several different geometries \cite{crow75,mora03}.
The first-principle numerical calculation of the steady-state profiles reached by the expansion of a plasma plume in a large domain is very challenging, due to the wide separation of scales involved.
For instance, the plasma Debye length can vary by several orders of magnitudes in few hundred meters along the expansion trajectory. Therefore, numerical simulations are usually constrained to either the 
near- or the far-field expansion, i.e., to a restricted simulation domain \cite{taccogna02,boyd02,garrigues02,taccogna04,boyd06,taccogna07,kronhaus12}.\\
A very appealing alternative to expensive numerical simulations is, of course, to seek for analytical solutions of the plume expansion.
This is often done by assuming that the plasma obeys simplified fluid equations \cite{parks79, korsun97,ashkenazy01, korsun04,gabdullin08,peraire06}.\\
In this paper we analyze and discuss the self-similar solutions presented in Refs. \cite{parks79, korsun97, ashkenazy01}. It is important to notice that
the self-similar approach used in such models is different from the standard procedure followed in similarity methods described, for instance, in classical textbooks \cite{bluman_book}.
In fact, the standard approach is based on the idea of reducing the dimensionality of a set of partial differential equations, through an appropriate variable transformation, 
that can be identified by exploiting the symmetries of the system. On the other hand, the models described in this paper retain their full-dimensionality, and are based on a variable transformation
constructed on the principle that one variable is constant along fluid streamlines.
Another crucial point to notice, is that the solutions presented in Refs. \cite{parks79, korsun97, ashkenazy01} are approximate solutions. As we will show, the self-similarity assumption 
is indeed inconsistent with the fluid model employed and, as such, the solutions obtained do not exactly solve the initial set of equations.
However, such solutions, albeit inexact, have the advantage that they can be quickly evaluated on an arbitrary large domain. For this reason, they can still be valuable, if one is able to 
estimate how large their error is with respect to the true solution, and, of course, if such error is reasonably small.\\
The aim of this paper is twofold. First, we analyze the earlier solutions of Refs. \cite{parks79, korsun97, ashkenazy01} and we measure their errors with respect to the numerical solution
of the underlying fluid equations. Second, we present a generalization of such solutions, in an attempt to provide a more flexible family of solutions that can have a wider 
range of applicability.\\
The paper is organized as follows. Section 1 presents the mathematical model, the self-similarity assumption, and the resulting set of equations.
In Section 2, we describe the approximate solutions presented in Refs. \cite{parks79,korsun97, ashkenazy01}, and we comment on their range of validity and respective errors.
In Section 3, we describe a new class of solutions, that is based on a generalization of previous models, and we show that they are indeed applicable to a wider class of situations.
In Section 4 we show how to easily extend the solutions that were derived for a plasma with no azimuthal velocity to a more general case.
Finally, we draw conclusions of this study in Section 5.

\section{Mathematical model}\label{model}
We study a system composed by collisionless, singly charged, ions and electrons, in a steady-state, axisymmetric configuration.
We assume that electrons inertia can be neglected, quasi-neutrality holds, the plasma is electrostatic, and there is no background magnetic field.
For simplicity, we assume cold ions although a polytropic equation of state for ions might be easily incorporated in the model.
By employing cylindrical coordinates $(r,z,\theta)$ the ion continuity equation and the conservation of ion and electron momentum read:
\begin{eqnarray}
\frac{\partial (nu_z)}{\partial z} +\frac{1}{r}\frac{\partial(rnu_r)}{\partial r}&=&0\label{continuity} \\ 
u_z\frac{\partial u_r}{\partial z}+u_r\frac{\partial u_r}{\partial r}-\frac{u_\theta^2}{r}&=&-\frac{\partial \phi}{\partial r}\label{mom_r}\\
u_z\frac{\partial u_z}{\partial z}+u_r\frac{\partial u_z}{\partial r}&=&-\frac{\partial \phi}{\partial z}\label{mom_z}\\
u_z\frac{\partial u_\theta}{\partial z}+ u_r\frac{\partial u_\theta}{\partial r}+\frac{u_\theta u_r}{r}&=&0\label{eq_theta}\\
\frac{1}{n}\nabla p_e&=&\nabla \phi\label{mom_el}\\
\end{eqnarray}
where the symmetry condition $\frac{\partial}{\partial\theta}=0$ has been used, $n$ is the density, $u_z$ and $u_r$ are the ion axial and radial velocities, 
$\phi$ is the electrostatic potential, and $p_e$ is the electron pressure.
Eqs. (\ref{continuity})-(\ref{mom_el}) are complemented by a polytropic equation of state for the electrons:
\begin{equation}
 p_e=n^\gamma
\end{equation}
with $\gamma$ the polytropic index. Quantities have been normalized as follows: velocities to $\sqrt{T_0/m_i}$ (with $T_0$ a reference electron temperature and $m_i$ the ion mass), the electrostatic potential
to $T_0/e$ (with $e$ the elementary charge), density to a reference density $n_0$, lengths to a characteristic length $R$, and pressure to $n_0T_0$.
In general, the value of the polytropic index $\gamma$ depends on the degree of ionization of the plasma \cite{burm99}. For instance, \cite{dannenmayer13} have reported experimental results for thrusters, with $\gamma$ lower than $5/3$, and \cite{taccogna14} have used the value 1.3 for their PIC simulations.
However, as shown in Ref.\cite{burm99}, $\gamma$ tends to the neutral gas theoretical value for a fully ionized plasma, which, for simplicity, is the case treated in this paper. Hence, for all the cases presented here $\gamma=5/3$, i.e. the adiabatic constant of the monoatomic perfect gas.\\
The model in Eqs. (\ref{continuity})-(\ref{mom_el}) must be interpreted as a boundary value problem in the two-dimensional plane $(r,z)$. Since the equations involved are first order 
partial differential equations, one is allowed to specify the boundary conditions at $r=0$ and $z=0$, which determine the solution in the whole domain.
Also, note that the electron pressure and the electric potential can be substituted in Eqs. (\ref{mom_r})-(\ref{mom_z}), and post-processed after the solution for $n$, $u_r$, $u_z$, and $u_\theta$ has been obtained.

\subsection{Self-similar solution}
We now elucidate the procedure to seek for a self-similar solution of Eqs.(\ref{continuity})-(\ref{mom_el}), extending the derivation in \cite{merino11} to the case $u_\theta\neq0$.
First, we introduce a change of variable $(r,z)\rightarrow(\eta,z)$, with $\eta=\frac{r}{a(z)}$, and $a(z)$ an unspecified function. Then, we assume that the unknowns $n,u_z,u_r,u_\theta$ can be factorized as:
\begin{eqnarray}
 n(\eta,z)&=&n_c(z)n_t(\eta)\label{eq_n}\\
 u_z(\eta,z)&=&u_c(z)u_t(\eta)\\
 u_r(\eta,z)&=&\eta a'(z)u_z(\eta,z),\label{eq_ur}\\
 u_\theta(r,\eta)&=&\frac{\eta}{a(z)}Eu_t(\eta),\label{eq_utheta} 
\end{eqnarray}
where prime indicates differentiation, and $E$ is an arbitrary constant.
The definition for $u_r$, Equation (\ref{eq_ur}), follows from assuming that lines of constant $\eta$ correspond to streamlines, that is
\begin{equation}
 \left.\frac{\partial \eta}{\partial r}\right|_z u_r + \left.\frac{\partial \eta}{\partial z}\right|_r u_z=0.
\end{equation}
As we anticipated, the set of equations (\ref{continuity})-(\ref{mom_el}) is actually inconsistent with the separation of variables assumed in (\ref{eq_n})-(\ref{eq_utheta}). That is, the only solution that satisfies such separation
of variables is the trivial solution with $n, u_z, u_r, u_\theta$ all constant. 
However, the separation of variables is still worth considering if one can derive a class of approximate solutions for our model that are much faster 
to compute than the exact numerical solution, and its error relative to the full solution is small.\\
It is straightforward to prove that Eq.(\ref{eq_theta}) is automatically satisfied by using the factorization assumption in Eqs.(\ref{eq_n})-(\ref{eq_utheta}).
Substituting Eqs.(\ref{eq_n})-(\ref{eq_ur}) into Eqs.(\ref{continuity})-(\ref{mom_el}) and separating each equation in $\eta$ and $z$ dependent terms, 
one can get the following set of equations for 
$u_c(z)$, $n_c(z)$, and $a(z)$:
\begin{eqnarray}
 a^2n_cu_c&=&A\label{z_cont}\\
 \frac{1}{2}u_c^2+\frac{\gamma}{\gamma-1}n_c^{\gamma-1}&=&B\label{z_momz}\\
 n_c&=&a^D\label{z_momz2}\\
 \frac{u_ca}{n_c^{\gamma-1}}\left(a'u_c\right)'-\frac{E^2}{a^2n_c^{\gamma-1}}&=&\gamma C\label{z_momr}
\end{eqnarray}
and the the following equations for $n_t(\eta)$ and $u_t(\eta)$:
\begin{eqnarray}
 n_t^{2-D(\gamma-1)}&=&u_t^{-2D}\label{r_momz}\\
 \frac{ n_t^{\gamma-2}n_t'}{\eta u_t^2}&=& -C,\label{r_momr}
\end{eqnarray}
where $A$, $B$, $C$ and $D$ are arbitrary (separation) constants. Details of the calculation can be found in the Appendix.
Note that Eq.(\ref{z_cont}) descends from the continuity equation (\ref{continuity}), Eqs. (\ref{z_momz}), (\ref{z_momz2}), (\ref{r_momz}) from the momentum equation in the $z$ direction (\ref{mom_z}),
and Eqs. (\ref{z_momr}) and (\ref{r_momr}) from the momentum equation in the $r$ direction (\ref{mom_r}). Clearly,
we have a set of 6 equations for only 5 unknowns, and the system is overdetermined.

\section{Approximate solutions for $u_\theta=0$}
In order to derive an approximate solution of the model, a common approach is to assume $u_c=const$ and to prescribe a certain profile 
for $u_t$. In this way one can still satisfy some (but not all) of the equations (\ref{z_cont})-(\ref{r_momr}).
For simplicity, and to be adherent to earlier literature, we consider in this section the case where $E=0$, and thus $u_\theta=0$.
All of the solutions discussed here satisfy the continuity equation and the conservation of momentum in $r$, but not in $z$. From equations (\ref{z_cont}), (\ref{z_momr}), and (\ref{r_momr}), it follows that
\begin{eqnarray}
 n_c(z)&=&\frac{a^2(0)n_c(0)}{a^2(z)}\label{eq_nc}\\
 a''(z)&=&a^{1-2\gamma}(z)\frac{\gamma C}{u_c^2}a(0)^{2(\gamma-1)}n_c(0)^{1-2\gamma}\label{eq_a}\\
 n_t^{\gamma-2}n_t'&=&-\eta u_t^2C.\label{eq_nt}
\end{eqnarray}
In addition to the value of $u_c$ and the profile of $u_t$, Eqs. (\ref{eq_nc})-(\ref{eq_nt}) require to specify the value of $C$, $a(0)$, $a'(0)$, and $n_c(0)$.
As we will see some of these free parameters depend on the boundary conditions at $z=0$, i.e., the boundary from which the plume is injected.
We assume the following boundary conditions:
\begin{eqnarray}
 n(r=0,z=0)&=&1\label{bc_n}\\
 n(r=R,z=0)&=&0.01\label{bc_n2}\\
 u_r(r=1,z=0)&=&1\label{bc_ur}\\
 \left.\frac{\partial n}{\partial r}\right|_{r=0}&=&0\label{bc_4}\\
 \left.\frac{\partial u_z}{\partial r}\right|_{r=0}&=&0\label{bc_5}\\
 u_r(r=0,z) &=& 0 \label{bc_6} .
\end{eqnarray}
Eqs. (\ref{bc_n}) and (\ref{bc_ur}) are simply normalization constraints. Eqs. (\ref{bc_n2}) requires that the profile of the density at injection decreases by a factor of 100 
within the characteristic length $R$. Eqs. (\ref{bc_4})-(\ref{bc_6}) are 
symmetry conditions at $r=0$.\\
In this section, we describe the solutions presented in \cite{parks79}, \cite{korsun97}, \cite{ashkenazy01}, and we propose a new class of solutions
that include earlier models as limiting cases and produce an higher degree of flexibility in terms of plume injection profiles that can be represented  at $z=0$.\\
Such approximate solutions provide the density and velocity profiles at the boundary $z=0$ which are used to calculate the (exact) numerical solution. The latter is evaluated on a domain
$[r,z]=[0,R]\times[0,80]$, with $R=50$ and grid size $\Delta r=\Delta z=0.2$. The derivatives in $r$ and $z$ are discretized
with a fourth order central (5-points stencil) and a third order upwind (4-points stencil) difference scheme, respectively.\\
Since we are only interested in the steady state solution, we have used the following marching scheme for solving numerically Eqs.(\ref{continuity})-(\ref{mom_el}). 
First, the solution is calculated on the partial domain $[0,R]\times[0,4\Delta z]$ with an iterative Newton-GMRES solver \cite{kelley}.
The solution is then extended to the full domain by adding one row of cells in $z$ at the time and calculating the new solution at each step. This is possible since we 
use an upwind discretization in $z$ that requires only the points where the solution has already been calculated.\\
For each approximate solution we evaluate the error with respect to the full numerical solution. We define the following two measures of error:
\begin{eqnarray}
 \varepsilon_r=\mathrm{max}\left|\frac{(n^au_r^a-n^nu_r^n)}{n^nu_r^n}\right|\cdot 100\\
 \varepsilon_z=\mathrm{max}\left|\frac{(n^au_z^a-n^nu_z^n)}{n^nu_z^n}\right|\cdot 100,
 \end{eqnarray}
which are the $L_1$ norm of the relative error of the radial and axial fluxes, where the superscripts $a$ and $n$ indicate the approximate and the numerical solutions, respectively.
Each of the solutions presented in Refs. \cite{parks79,korsun97,ashkenazy01} makes a different assumption on $u_t$, and we proceed to discuss them separately.

\subsection{Parks and Katz (PK) solution}
\begin{figure}
\noindent\includegraphics[width=20pc]{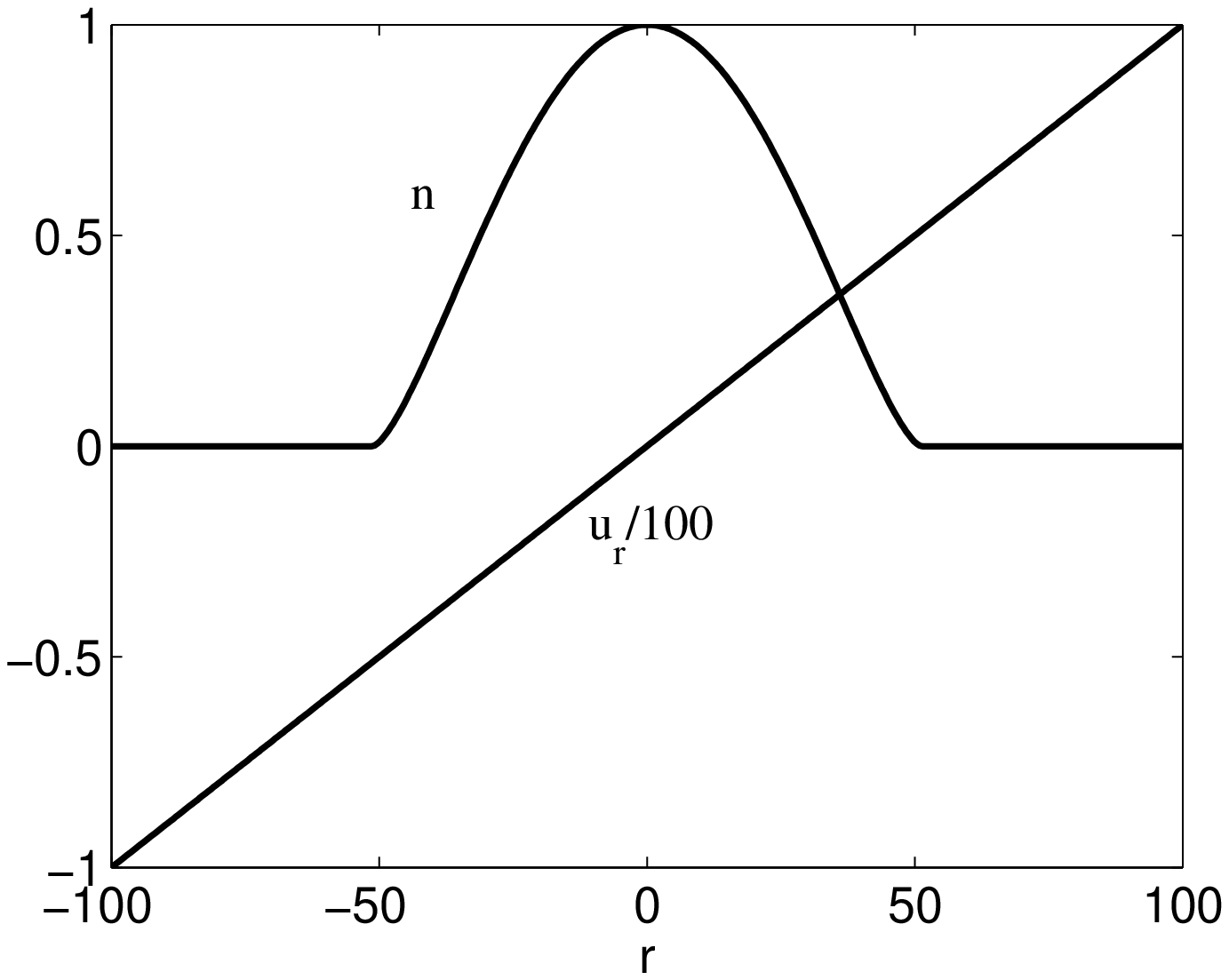}
 \caption{PK solution. Profiles for the density $n$ and the radial velocity $u_r$ (rescaled by 100) at the injection boundary $z=0$. In the region $r\gtrsim 50$ where the density becomes
 imaginary, it is artificially set equal to zero.}\label{fig:profile_PK}
\end{figure}

\begin{figure}
\noindent\includegraphics[width=20pc]{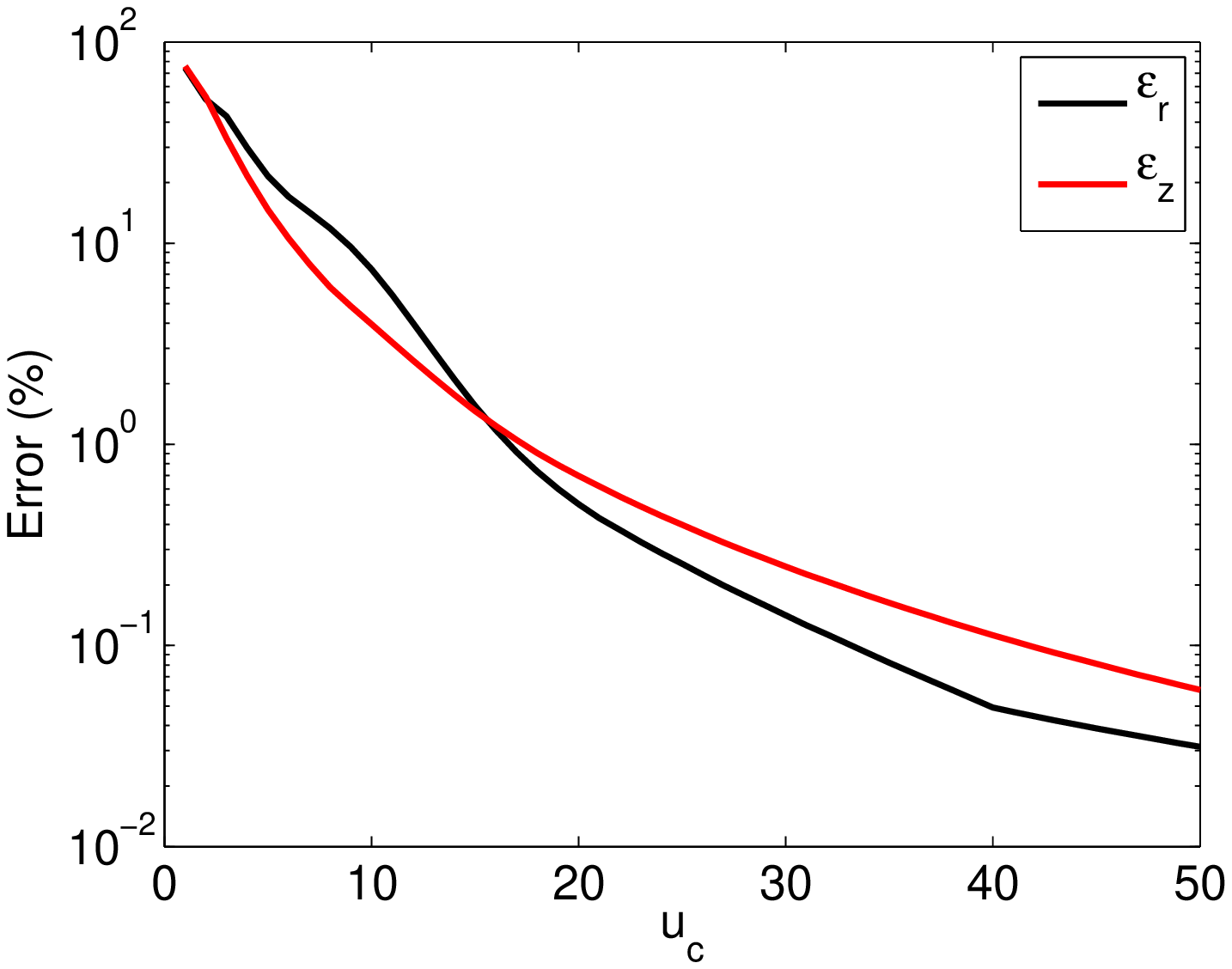}
 \caption{PK solution. Errors $\varepsilon_r$ (black) and $\varepsilon_z$ (red) as function of $u_c$.}\label{fig:error_PK}
\end{figure}

The solution proposed in \cite{parks79} assumes a constant $u_t=1$. Solving Eq.(\ref{eq_nt}) for $n_t$ gives the profile:
\begin{equation}
 n_t=\left(C_0-\frac{\gamma-1}{2}C\eta^2\right)^{\frac{1}{\gamma-1}},
\end{equation}
where $C_0$ is an integration constant. Following \cite{merino11} we set $C_0=1$, therefore the PK solution reads:
\begin{eqnarray}
 n(r,z)&=& \frac{a^2(0)}{a^2(z)}\left(1-\frac{\gamma-1}{2}C\frac{r^2}{a^2(z)}\right)^{\frac{1}{\gamma-1}}\label{PK_n}\\
 u_r(r,z)&=&\frac{r}{a(z)}a'(z)u_c\label{PK_ur}\\
 u_z(r,z)&=&u_c\label{PK_uz},
\end{eqnarray}
where we have used $n_c(0)=1$, imposed by Eq. (\ref{bc_n}). The boundary condition (\ref{bc_n2}) allows to solve for $C$ and Eq. (\ref{bc_ur}) yields the relationship
\begin{equation}
 a(0)=u_ca'(0)
\end{equation}
In summary, the only two free parameters are $u_c$ and $a'(0)$ and the profile for $a(z)$ can be calculated by solving numerically Eq.(\ref{eq_a}).
Note that the profiles of $u_r$ and $n$ at the injection boundary $z=0$ do not depend on the particular choice of $u_c$ and $a'(0)$.
Figure \ref{fig:profile_PK} shows the profiles for the density and the radial velocity at $z=0$ as a function of $r$.
Clearly, the radial velocity increases linearly with $r$.
An important feature of the PK solution is that, for any given value of $z$, the solution is constrained to a certain range in $r$. 
Indeed, one can see from Eq.(\ref{PK_n}) that the density becomes a complex quantity (i.e. unphysical) for a large enough $r$. 
In Figure \ref{fig:profile_PK} we have artificially set $n=0$ in the regions where the density becomes imaginary.
Figure \ref{fig:density_PK} shows the density profile in the $(z,r)$ plane, for different values of the axial velocity $u_z=u_c$, in logarithmic scale, for a fixed value
of $a'(0)=0.2$. Once again, notice that the density profile at $z=0$ is not a function of $u_c$.
The errors $\varepsilon_r$ and $\varepsilon_z$ as a function of $u_c$ are shown in Figure \ref{fig:error_PK}. 
For $u_c>20$, $\varepsilon_r$ and $\varepsilon_z$ are smaller than $1\%$.
Note, however, that here and in the next similar figures, the error has been calculated in the restricted region $|r|\leq R$. The region where the density is artificially set equal to zero  is excluded by the evaluation of the error. The reason is that the marching routine described above is not very robust when a sharp gradient in the density is included as
boundary condition and, therefore, it is not straightforward to obtain a numerical solution that extends to $|r|>R$.

\subsection{Ashkenazy and Fruchtman (AF) solution}
\begin{figure}
\noindent\includegraphics[width=20pc]{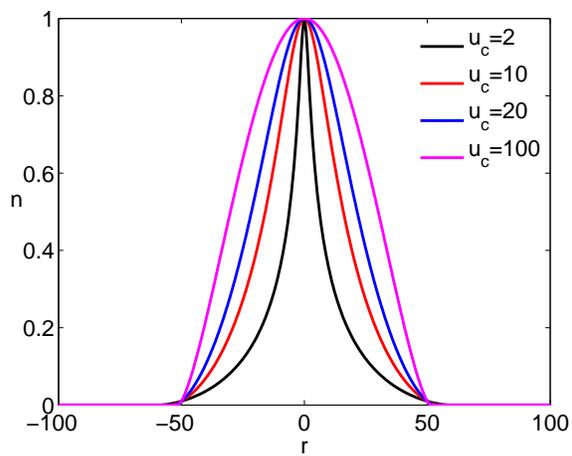}
 \caption{AF solution. Profile of the density $n$ at the injection boundary $z=0$, for $u_c=2, 10, 20, 100$ (in black, red, blue, and magenta, respectively). 
 In the region $r\gtrsim 50$ where the density becomes imaginary, it is artificially set equal to zero.}\label{fig:profile_AF_n}
\end{figure}

\begin{figure}
\noindent\includegraphics[width=20pc]{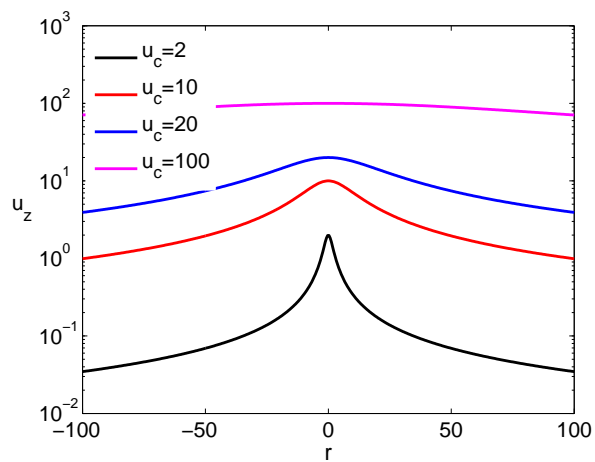}
 \caption{AF solution. Profile of the axial velocity $u_z$ at the injection boundary $z=0$, for $u_c=2, 10, 20, 100$ (in black, red, blue, and magenta, respectively).
 Vertical axis in logarithmic scale.
 }\label{fig:profile_AF_uz}
\end{figure}

\begin{figure}
\noindent\includegraphics[width=20pc]{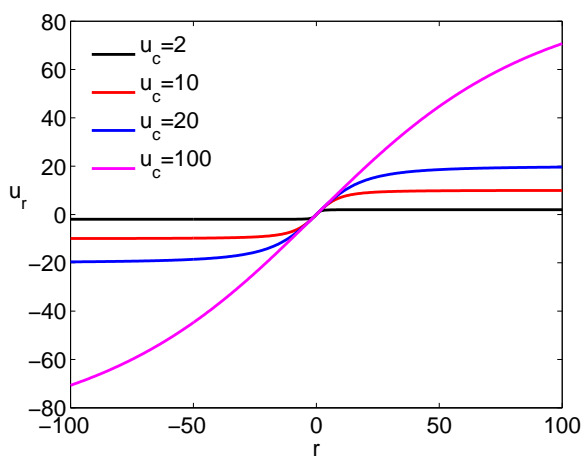}
 \caption{AF solution. Profile of the radial velocity $u_r$ at the injection boundary $z=0$, for $u_c=2, 10, 20, 100$ (in black, red, blue, and magenta, respectively). }\label{fig:profile_AF_ur}
\end{figure}

\begin{figure}
\noindent\includegraphics[width=20pc]{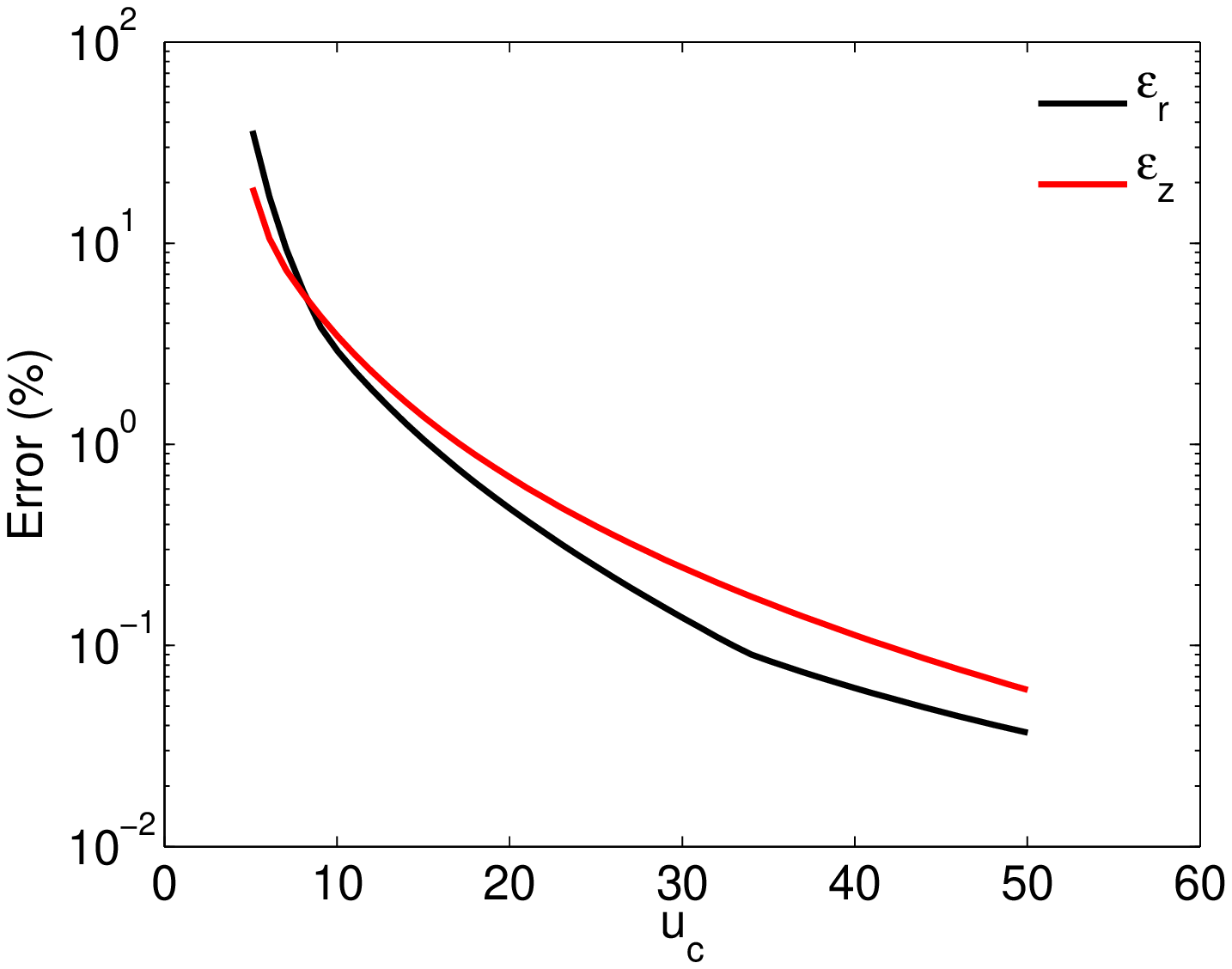}
 \caption{AF solution. Errors $\varepsilon_r$ (black) and $\varepsilon_z$ (red) as function of $u_c$.}\label{fig:error_AF}
\end{figure}

The AF solution \cite{ashkenazy01} assumes a conical velocity profile at the injection, such that 
\begin{equation}
 u_t=\left(1+a'(0)^2\eta^2\right)^{-\frac{1}{2}}
\end{equation}
The profile for $n_t$ is given by solving  Eq.(\ref{eq_nt}):
\begin{equation}
 n_t=\left(C_0-\frac{C(\gamma-1)}{2(a'(0))^2}\log\left(1+(a'(0))^2\eta^2\right)\right)^{\frac{1}{\gamma-1}},
\end{equation}
where $C_0$ is an integration constant that we set equal to 1.
The boundary condition (\ref{bc_n}) imposes $n_c(0)=1$, while Eq.(\ref{bc_n2}) yields an equation that must be solved to obtain the value of $C$.
The remaining free parameters $u_c$, $a(0)$, $a'(0)$ obey, via the boundary condition (\ref{bc_ur}), the relationship:
\begin{equation}
 \frac{a(0)}{a'(0)}\sqrt{1+\left(\frac{a'(0)}{a(0)}\right)^2}=u_c\label{AF_azero}
\end{equation}
For simplicity, we fix $a'(0)=0.2$, allowing $u_c$ to vary, and calculate $a(0)$ through Eq. (\ref{AF_azero}). As usual, $a(z)$ is calculated via Eq.(\ref{eq_a}).
In summary, the profiles for $n$, $u_r$, and $u_z$ are:
\begin{eqnarray}
 n(r,z)&=& \left(\frac{a(0)}{a(z)}\right)^2\left[1-\frac{C(\gamma-1)}{2(a'(0))^2}\log\left(1+\left(\frac{a'(0)}{a(z)}r\right)^2\right)\right]^{\frac{1}{\gamma-1}}\label{AF_n}\\
 u_r(r,z)&=&\frac{r}{a(z)}a'(z)u_c\left[1-\frac{C(\gamma-1)}{2(a'(0))^2}\log\left(1+\left(\frac{a'(0)}{a(z)}r\right)^2\right)\right]^{\frac{1}{\gamma-1}}\label{AF_ur}\\
 u_z(r,z)&=& \frac{u_c}{\sqrt{1+\left(\frac{a'(0)}{a(z)}r\right)^2}}\label{AF_uz}
\end{eqnarray}
Figure \ref{fig:profile_AF_n} shows the density profile for the AF solution at the injection boundary $z=0$, as
function of $r$, for the values $u_c=2$, 10, 20, 100. Similarly to the previous model, the density becomes imaginary for $r$ larger than certain values.
The corresponding profiles for $u_z$ and $u_r$ are plotted in Figures \ref{fig:profile_AF_uz} and \ref{fig:profile_AF_ur}, respectively.
Figure \ref{fig:density_AF} shows the two-dimensional density profile in the $(z,r)$ plane, in logarithmic scale for four values of $u_c$.
One can notice a qualitative similarity with the profiles of the PK solution, especially for large $u_c$. 
The errors for the AF solution are shown in Figure \ref{fig:error_AF} and are comparable with those obtained for the PK solution (Figure \ref{fig:error_PK}),
i.e., the errors are smaller than $1\%$, for $u_c\gtrsim20$.

\subsection{Korsun and Tverdokhlebova (KT) solution}
\begin{figure}
\noindent\includegraphics[width=20pc]{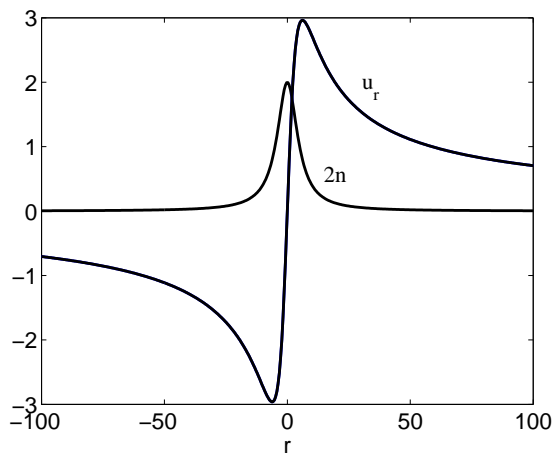}
 \caption{KT solution. Profiles for the density $n$ (rescaled by 2) and the radial velocity $u_r$ at the injection boundary $z=0$.}\label{fig:profile_KT}
\end{figure}

\begin{figure}
\noindent\includegraphics[width=20pc]{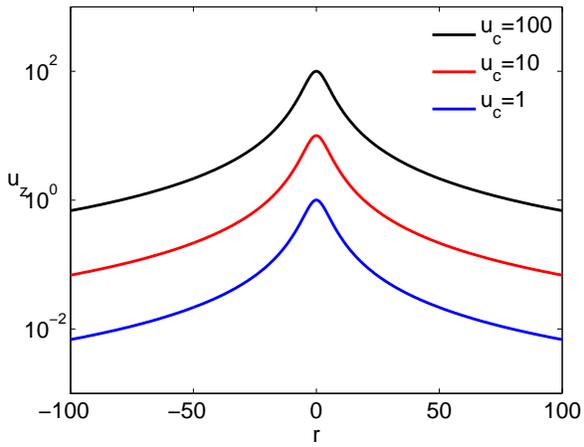}
 \caption{KT solution. Profile for the axial velocity $u_z$ at the injection boundary $z=0$, for $u_c=100, 10, 1$ (in black, red, and blue, respectively). The vertical axis 
 is in logarithmic scale.}\label{fig:profile_KT_uz}
\end{figure}

\begin{figure}
\noindent\includegraphics[width=40pc]{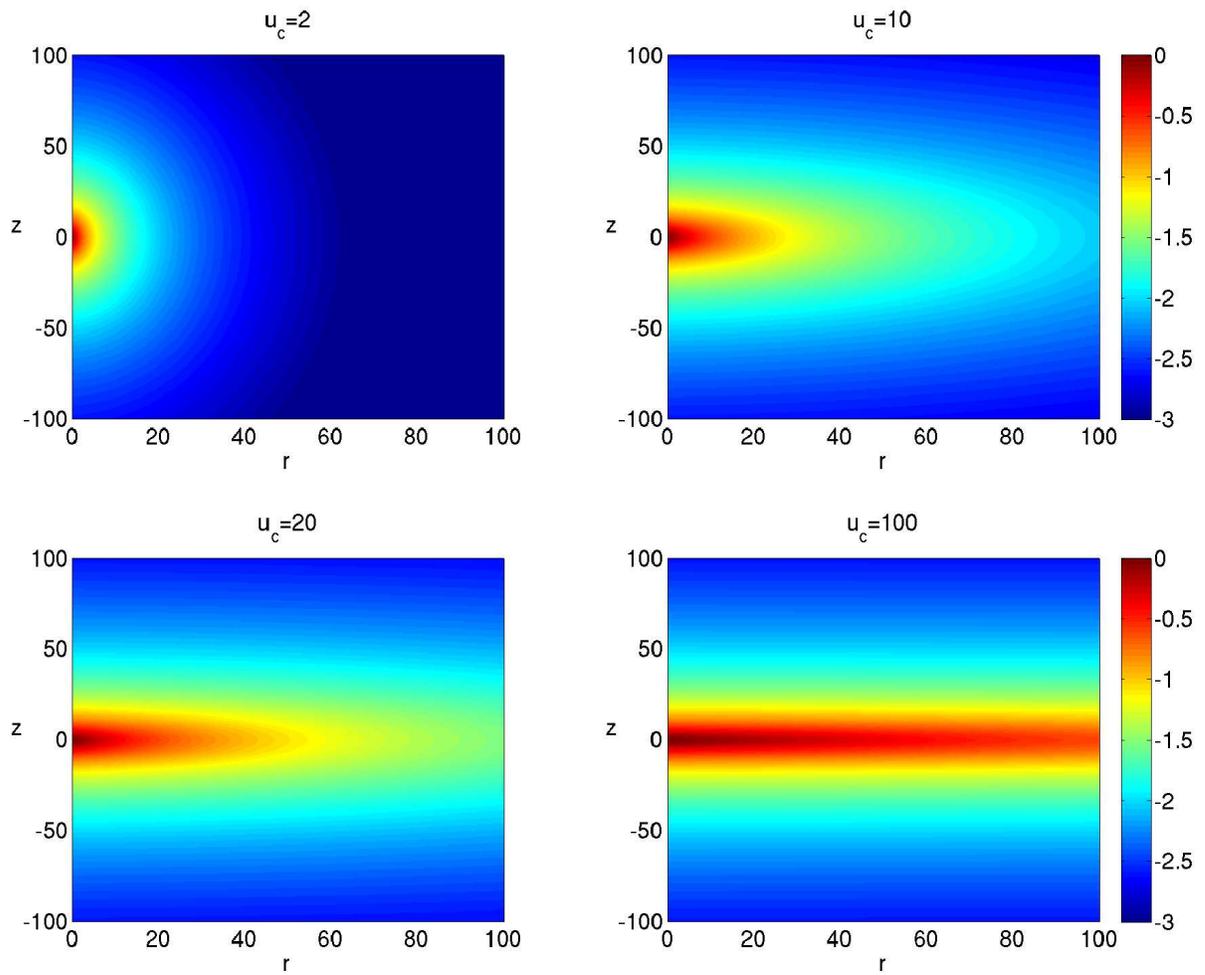}
 \caption{KT solution. Two-dimensional profiles of the density in $(z,r)$, for $u_c=2, 10, 20, 100$, in logarithmic scale. }\label{fig:density_KT}
\end{figure}

\begin{figure}
\noindent\includegraphics[width=20pc]{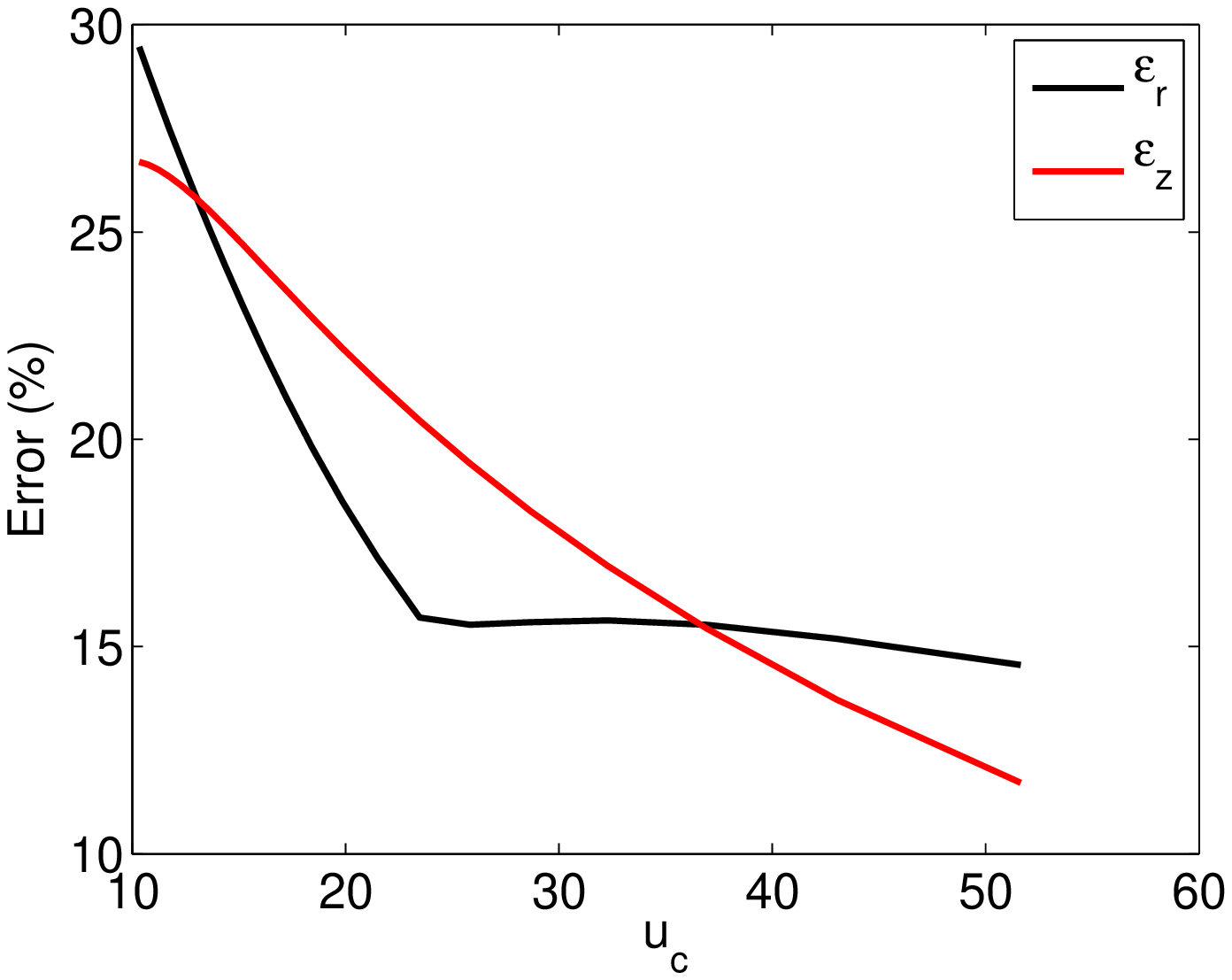}
 \caption{PK solution. Errors $\varepsilon_r$ (black) and $\varepsilon_z$ (red) as function of $u_c$.}\label{fig:error_KT}
\end{figure}

The solution proposed in \cite{korsun97} assumes the profile
\begin{equation}
 u_t(\eta)=\left(1-\frac{C}{2}\eta^2\right)^{-\frac{\gamma}{2}}
\end{equation}
from which Eq.(\ref{eq_nt}) dictates
\begin{equation}
 n_t(\eta)=\left(1-\frac{C}{2}\eta^2\right)^{-1}.
\end{equation}
Moreover, by setting $D=-2$, the KT solution satisfies Equation (\ref{r_momz}). By applying the boundary condition (\ref{bc_n}) and Eq. (\ref{z_momz2}), it follows that $a(0)=n_c(0)=1$. 
In summary, the KT solution not only satisfies the equations of continuity and momentum in $r$, but also Eqs.(\ref{z_momz2}) and (\ref{r_momz}), 
i.e., it satisfies 2 out of 3 equations for the conservation of momentum in $z$ direction.
The profiles for $n$, $u_r$, and $u_z$ are:
\begin{eqnarray}
 n(r,z)&=&\frac{1}{a^2(z)}\left(1-\frac{C}{2}\frac{r^2}{a^2(z)}\right)^{-1}\label{KT_n}\\
 u_r(r,z)&=&\frac{r}{a(z)}a'(z)u_c\left(1-\frac{C}{2}\frac{r^2}{a^2(z)}\right)^{-\frac{\gamma}{2}}\label{KT_ur}\\
 u_z(r,z)&=&u_c\left(1-\frac{C}{2}\frac{r^2}{a^2(z)}\right)^{-\frac{\gamma}{2}}\label{KT_uz}.
\end{eqnarray}
The boundary condition (\ref{bc_n2}) provides the expression for $C=2\left(1-\frac{1}{0.01}\right)/R^2=-0.0792$,
and $a(z)$ is calculated via Eq.(\ref{eq_a}). Hence, the only free parameter is $u_c$ which is related to $a'(0)$ through the relationship
\begin{equation}
 a'(0)=\frac{1}{u_c}\left(1-\frac{C}{2}\right)^{\frac{\gamma}{2}},
\end{equation}
that is obtained by applying Eq.(\ref{bc_ur}).\\
Similarly to the PK solution, the density and radial velocity profiles at the injection boundary $z=0$ are not a function of $u_c$. They are shown in Figure \ref{fig:profile_KT}.
The profile for $u_z$ is a linear function of $u_c$. Three examples for the values $u_c=1$, 10, 100 are shown in Figure \ref{fig:profile_KT_uz}.
Four examples of two-dimensional density profiles are shown in logarithmic scale in Figure \ref{fig:density_KT}, for $u_c=2$, 10, 20, 100.
The errors  $\varepsilon_r$ and $\varepsilon_z$ as a function of $u_c$ are shown in Figure \ref{fig:error_KT}. Notice that they are much higher than the respective errors for the 
PK and AF solutions. This might be surprising since the KT solution satisfies equations (\ref{z_momz2}) and (\ref{r_momz}), while the PK and AF solutions do not. 
However, given the nonlinear nature of the model, one cannot predict how large the errors will be with respect to the true solution. Moreover, as discussed in the Appendix, Eqs. (\ref{z_momz2}) and (\ref{r_momz})
follow from the assumption that Eq. (\ref{z_momz}) holds. Hence, satisfying equations (\ref{z_momz2}) and (\ref{r_momz}) does not automatically yields a lower error, since Eq. (\ref{z_momz}) is still not satisfied.

\subsection{Discussion on the PK, AF, and KT solutions}
\begin{figure}
\noindent\includegraphics[width=40pc]{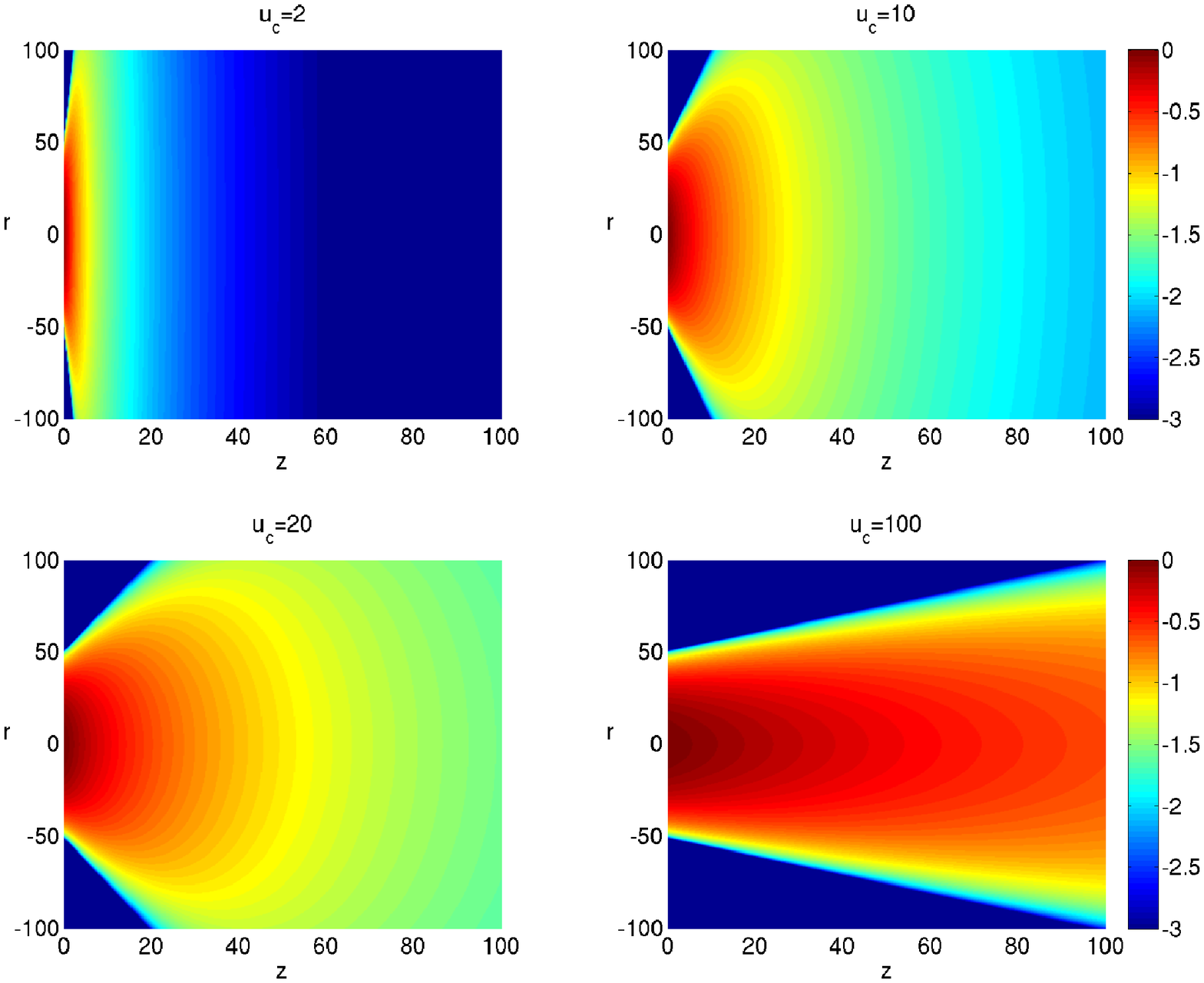}
 \caption{PK solution. Two-dimensional profiles of the density in $(z,r)$, for $u_c=2, 10, 20, 100$, in logarithmic scale. The density is artificially set equal to zero in 
 the regions where it becomes imaginary.}\label{fig:density_PK}
\end{figure}

\begin{figure}
\noindent\includegraphics[width=40pc]{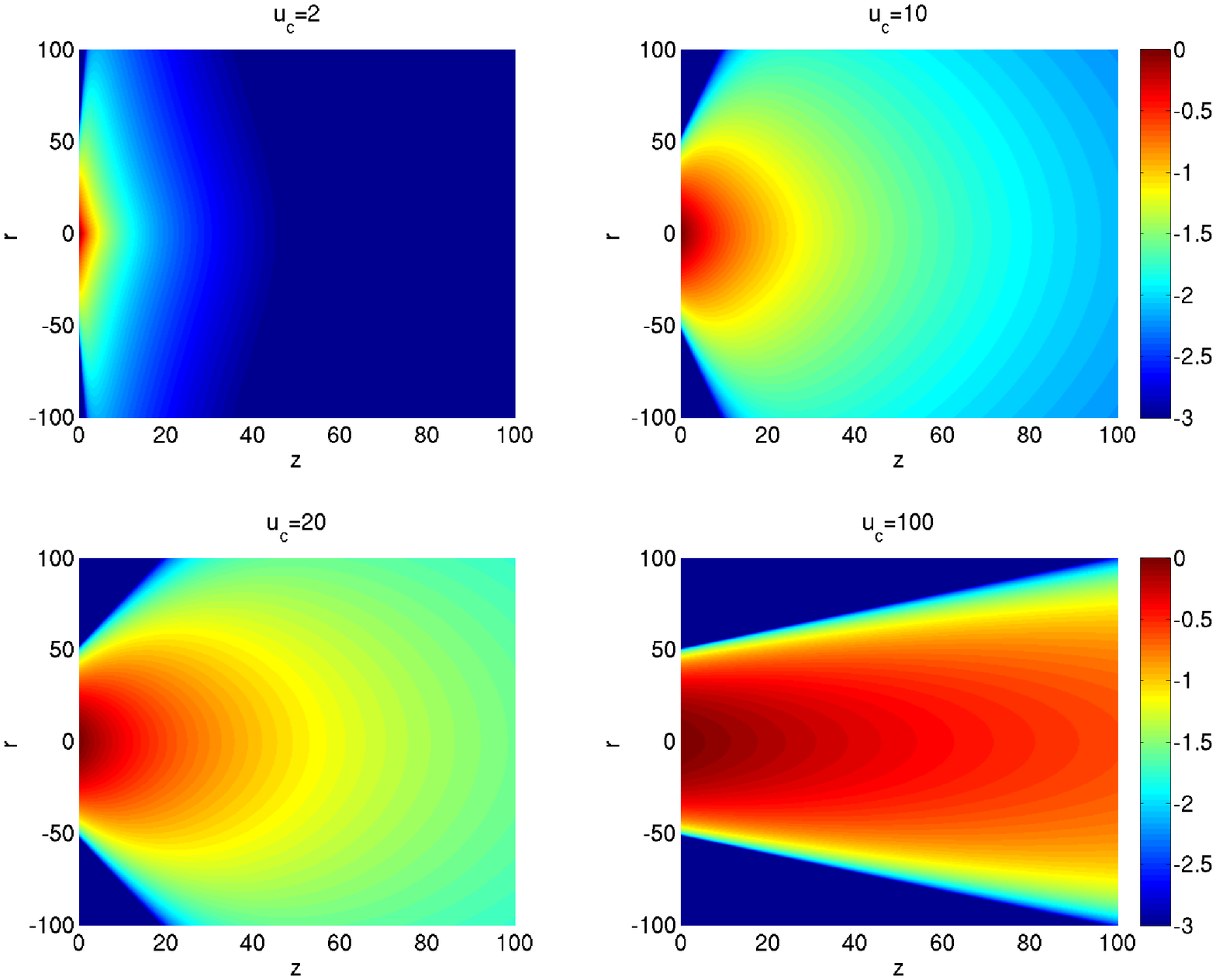}
 \caption{AF solution. Two-dimensional profiles of the density in $(z,r)$, for $u_c=2, 10, 20, 100$, in logarithmic scale. The density is artificially set equal to zero in 
 the regions where it becomes imaginary.}\label{fig:density_AF}
\end{figure}

\begin{figure}
\noindent\includegraphics[width=20pc]{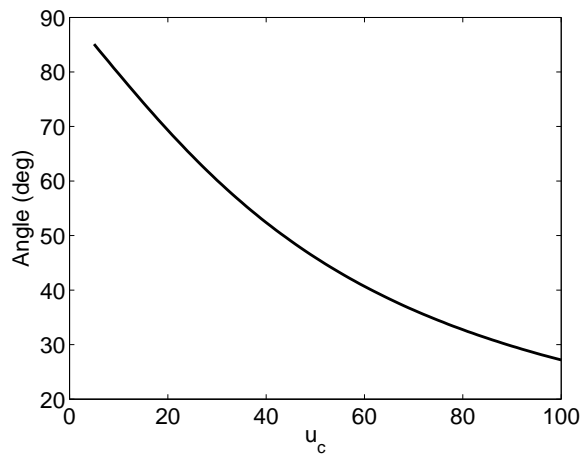}
 \caption{PK and AF solutions. Angle (in degrees), with respect to the axis $r=0$, that determines the (approximately straight) isocontour $n=0$, as a function of 
 the injection velocity $u_c$. The expansion is bounded within this cone angle.}\label{fig:angle}
\end{figure}

We have noted a certain similarity between the PK and AF solutions for large $u_c$. 
This is not surprising if one realizes that the AF solution tends exactly to the PK solution 
in the limit $u_c\rightarrow\infty$. Indeed, in this limit, Eq. (\ref{AF_azero}) requires
that $a(0)\rightarrow\infty$ and/or $a'(0)\rightarrow 0$. Taylor expanding the logarithmic term in $n$ and $u_r$
[Eqs. (\ref{AF_n})-(\ref{AF_ur})] one has 
\begin{equation}
 \log\left(1+\left(\frac{a'(0)}{a(z)}r\right)^2\right)\rightarrow \left(\frac{a'(0)r}{a(z)}\right)^2,
\end{equation}
and Eqs. (\ref{AF_n})-(\ref{AF_uz}) reduce exactly to Eqs. (\ref{PK_n})-(\ref{PK_uz}).\\
Although the PK and AF solutions yield relatively good results in terms of small errors with respect
to the numerical solutions (especially for large $u_c$), they are severely limited in the choice of density and velocity profiles at the injection.
In fact, the value of $u_c$ determines, by construction, the injection profiles that, in turn, determine the region in which the plume propagate.
From Figures \ref{fig:density_PK} and \ref{fig:density_AF}, it is evident that the cone of propagation becomes more and more collimated as $u_c$ increases.
In Figure \ref{fig:angle}, we show the angle (in degrees) of the isocontour $n=0$ with respect to the axis $r=0$, as a function of $u_c$.
Note that the cone of propagation is identical for the AF and PK solutions. Also, although the isocontour $n=0$ is not exactly a straight line in the $(z,r)$ plane,
it can be considered approximately straight in the region plotted in Figures \ref{fig:density_PK} and \ref{fig:density_AF}.
We emphasize that this cone of propagation, and the relative angles shown in Figure \ref{fig:angle} are an intrinsic characteristic of the PK and AF solutions,
which depends solely on the value of $u_c$.
On the other hand, the user of such approximate solutions might need to be able to choose a different injection profile than the one imposed by these solutions, since these depend on the source generating the contactor plasma.
In particular, as highlighted in Figure \ref{fig:angle}, large injection angles are not possible for a large injection velocity, for the PK and AF solutions.
Differently from PK and AF, the KT solution does not involve a propagation cone. Indeed, from Equation (\ref{KT_n}), it is easy to see 
that the density tends to zero only asymptotically for $r\rightarrow\infty$. Unfortunately, although the KT profiles offer an alternative to the PK and AF solutions, they yield a much 
larger error and therefore are not preferable.\\
In the following section we introduce a new class of solutions that allows a rather wide choice of injection profiles, still yielding relatively small errors
with respect to the numerical solutions.

\section{A new class of solutions}
\begin{figure}
\noindent\includegraphics[width=20pc]{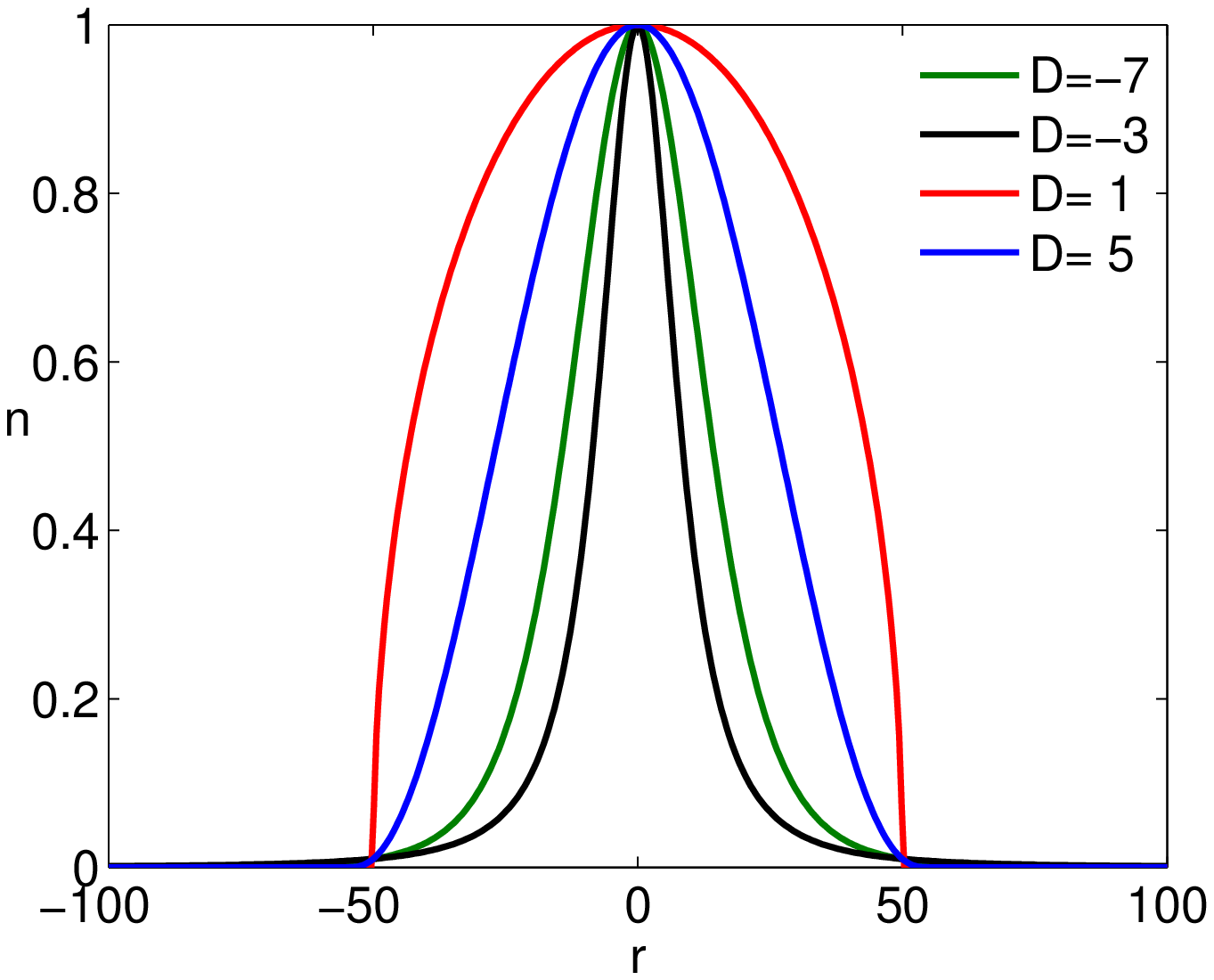}
 \caption{Generalized solution. Density profile at the injection boundary $z=0$ for $D=-7$, -3, 1, 5 (in green, black, red, and blue, respectively).}\label{fig:profile_new_n}
\end{figure}

\begin{figure}
\noindent\includegraphics[width=20pc]{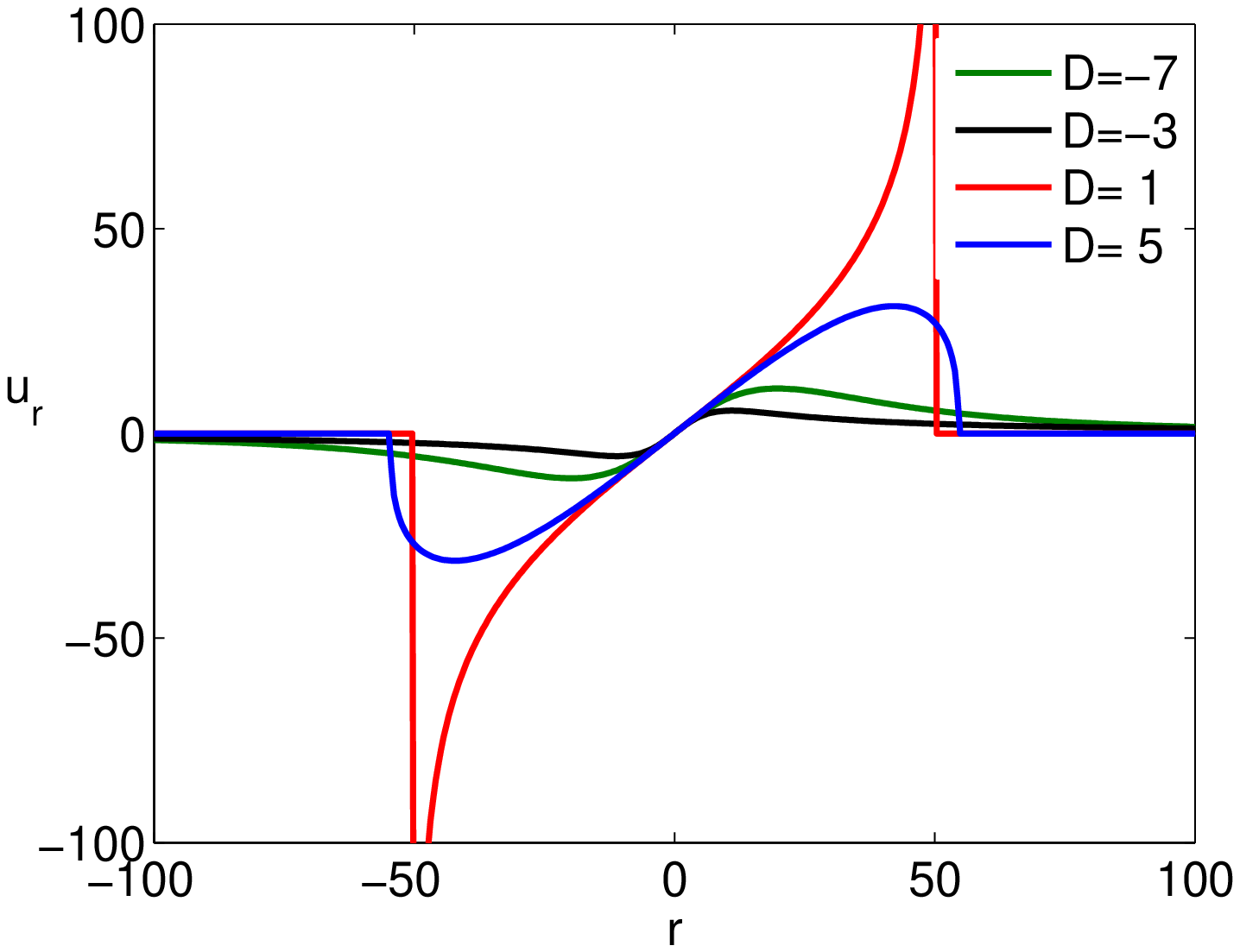}
 \caption{Generalized solution. Radial velocity profile at the injection boundary $z=0$ for $D=-7$, -3, 1, 5 (in green, black, red, and blue, respectively).}\label{fig:profile_new_ur}
\end{figure}

\begin{figure}
\noindent\includegraphics[width=20pc]{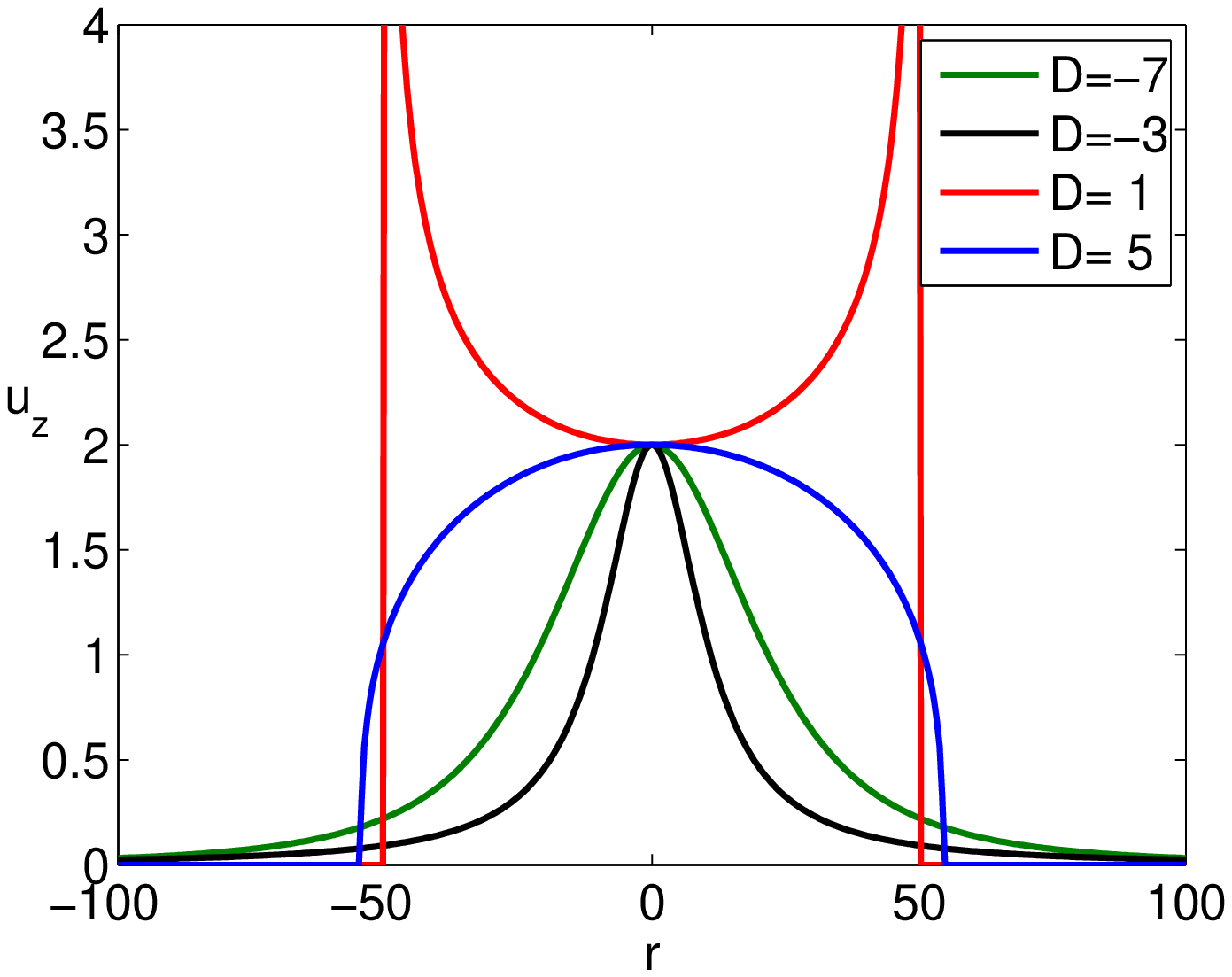}
 \caption{Generalized solution. Axial velocity profile at the injection boundary $z=0$ for $D=-7$, -3, 1, 5 (in green, black, red, and blue, respectively).}\label{fig:profile_new_uz}
\end{figure}

\begin{figure}
\noindent\includegraphics[width=40pc]{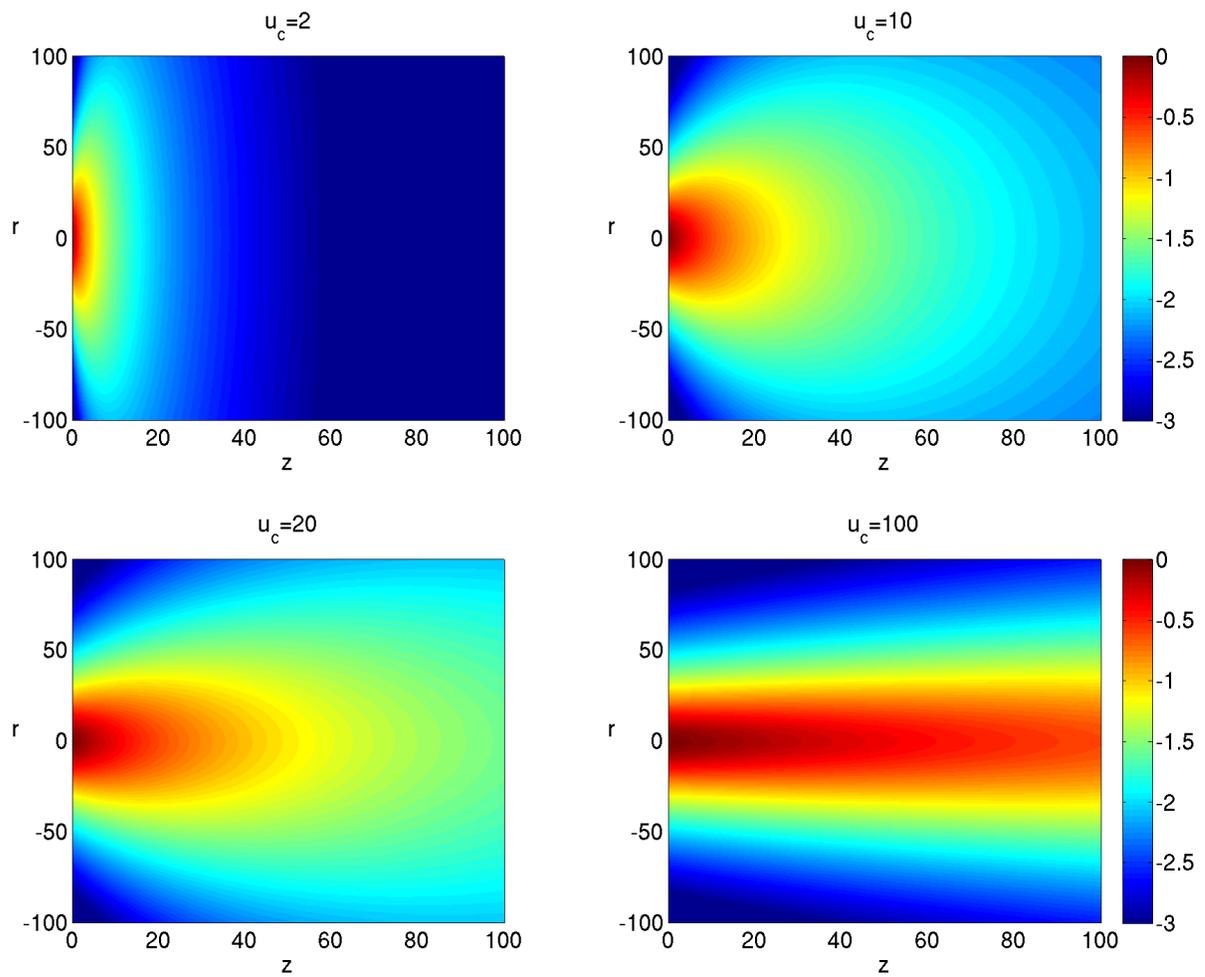}
 \caption{Generalized solution. Two-dimensional profiles of the density in $(z,r)$, for $u_c=2, 10, 20, 100$, and $D=-5$, in logarithmic scale. }\label{fig:profile_new_D_5_F_1}
\end{figure}

\begin{figure}
\noindent\includegraphics[width=40pc]{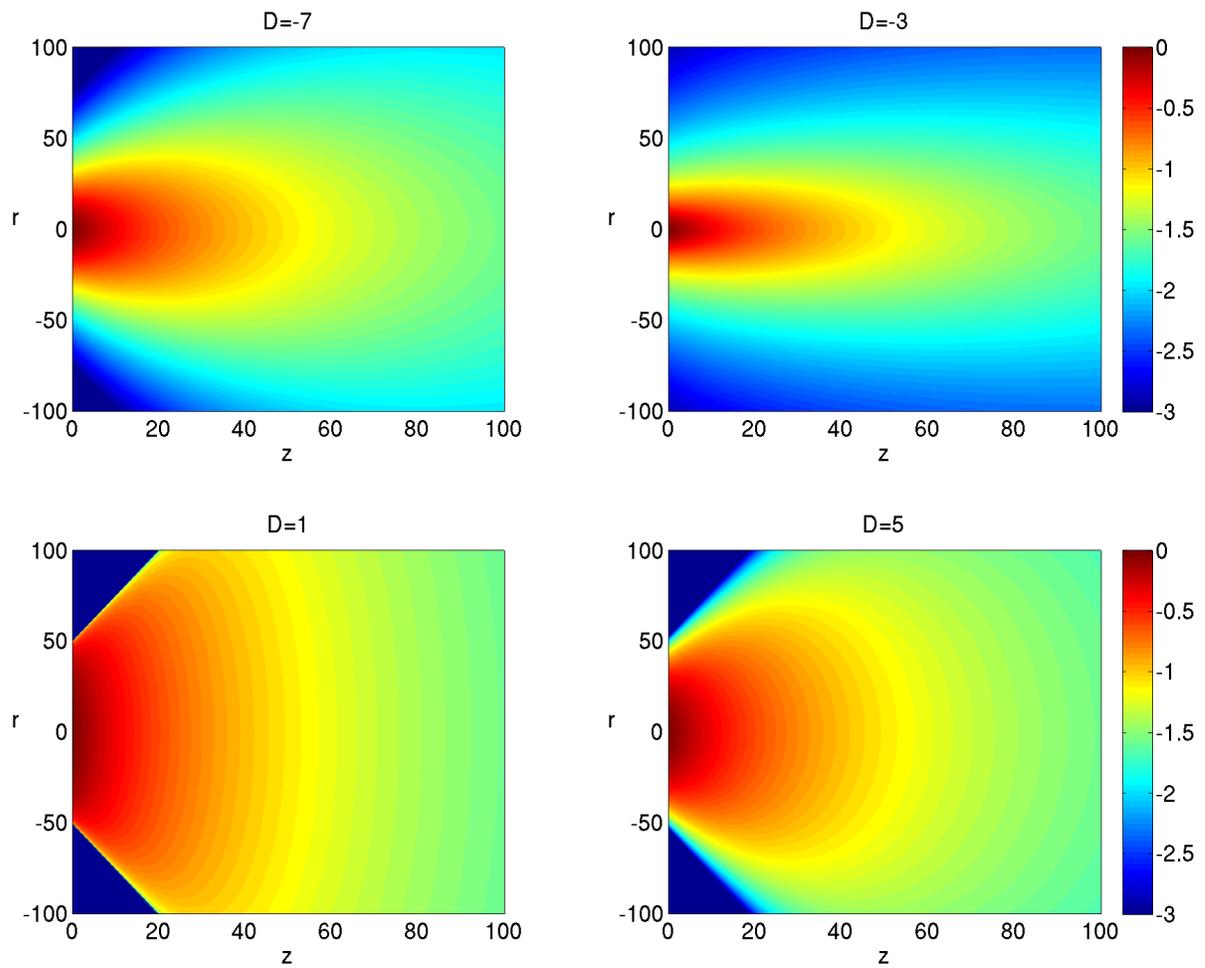}
 \caption{Generalized solution. Two-dimensional profiles of the density in $(z,r)$, for $D=-7$, -3, 1, 5, and $u_c=20$, in logarithmic scale. }\label{fig:profile_new_D_vary}
\end{figure}

\begin{figure}
\noindent\includegraphics[width=20pc]{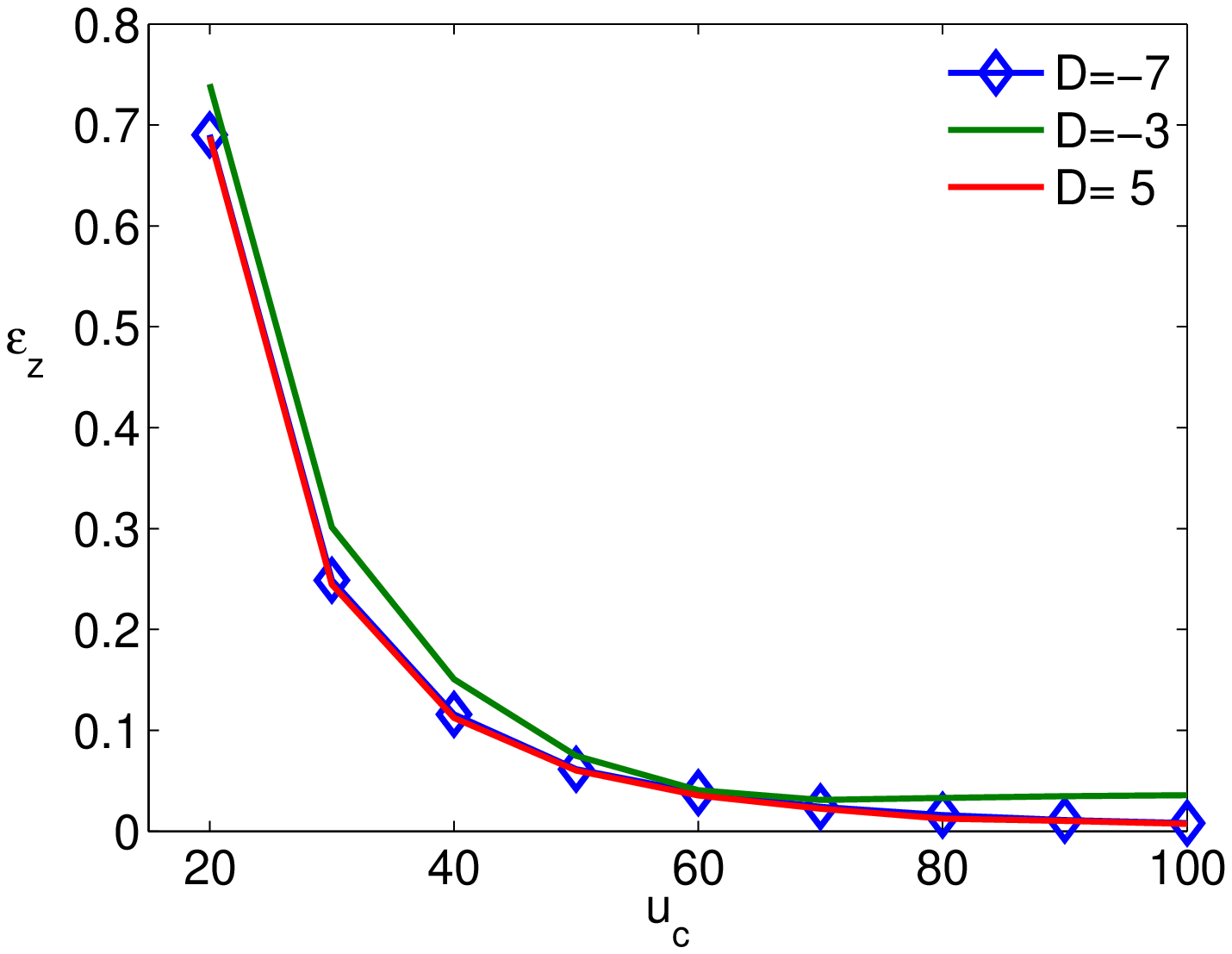}
 \caption{Generalized solution. Error $\varepsilon_z$ as a function of $u_C$ for $D=-7$, -3, 5 (in blue, green, red, respectively), and $F=1$.}\label{fig:error_new_z}
\end{figure}

\begin{figure}
\noindent\includegraphics[width=20pc]{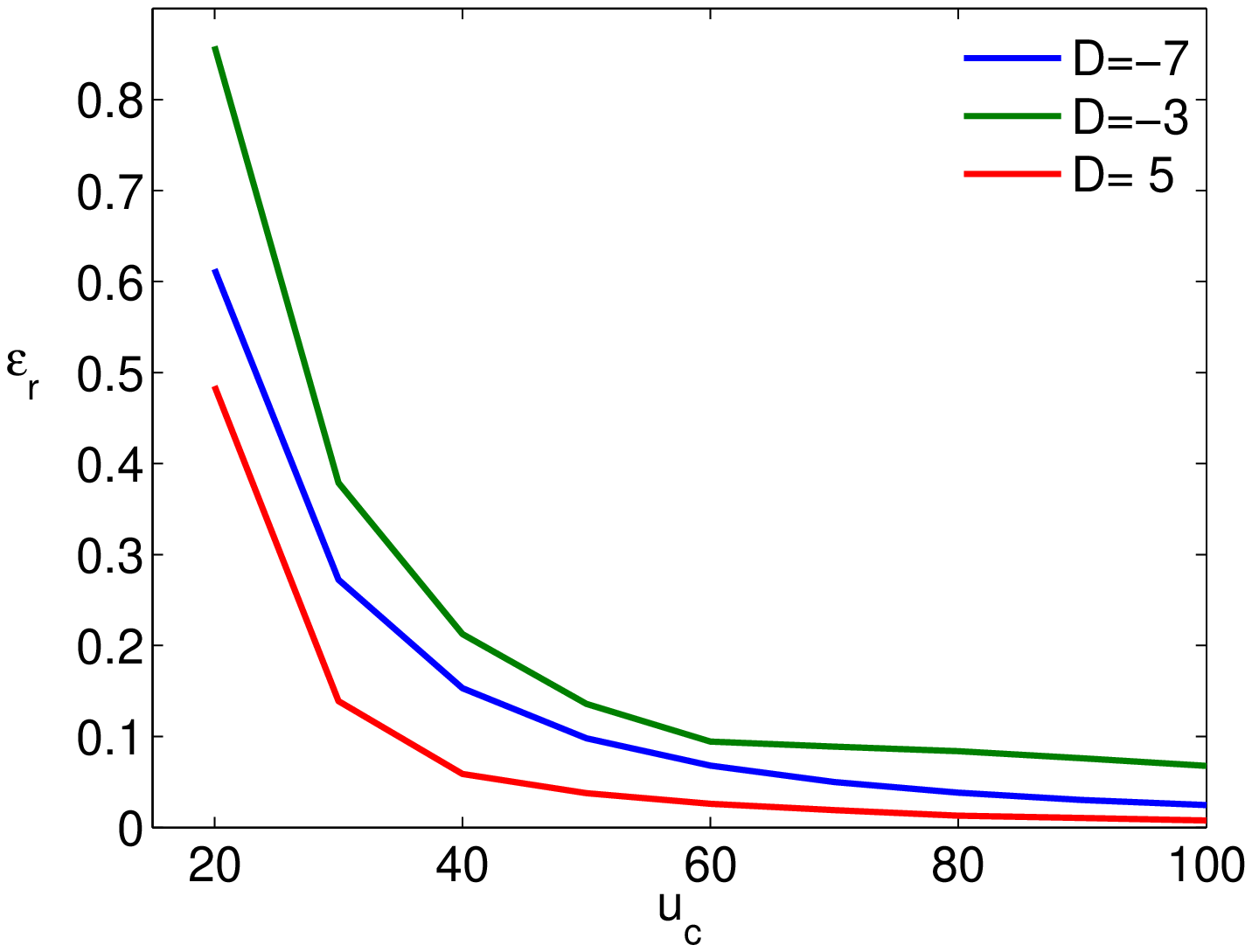}
 \caption{Generalized solution. Error $\varepsilon_r$ as a function of $u_C$ for $D=-7$, -3, 5 (in blue, green, red, respectively), and $F=1$.}\label{fig:error_new_r}
\end{figure}

The new class of solutions is based on a generalization of the KT and PK solutions. Interestingly, their error is much smaller than the errors yield for the KT solution, and comparable
to the ones of the PK and AF solutions.
As in previous models, we still assume a constant $u_c$. We also assume the following profiles for $n_t$ and $u_t$:
\begin{eqnarray}
 n_t(\eta)&=&\left(F-\frac{C}{D}\eta^2\right)^{\frac{D}{2}}\label{new_nt}\\
 u_t(\eta)&=& \left(F-\frac{C}{D}\eta^2\right)^{\frac{D(\gamma-1)}{4}-\frac{1}{2}}\label{new_ut}
\end{eqnarray}
Note that the KT solution belongs to this class of solution, for the particular case $D=-2$ and $F=1$, while for $D=2/(\gamma-1)$ and $F=1$, one recovers the PK solution.
Also, one can verify that Eqs.(\ref{new_nt})-(\ref{new_ut}) satisfy Eq. (\ref{r_momz}). \\
The solution reads:
\begin{eqnarray}
 n(r,z)&=&\frac{a^2(0)n_c(0)}{a^2(z)}\left(F+\frac{C}{D}\frac{r^2}{a^2(z)}\right)^{\frac{D}{2}}\label{new_n}\\
 u_r(r,z)&=&\frac{r}{a(z)}a'(z)u_c\left(F+\frac{C}{D}\frac{r^2}{a^2(z)}\right)^{\frac{D(\gamma-1)}{4}-\frac{1}{2}}\label{new_ur}\\
 u_z(r,z)&=&u_c\left(F+\frac{C}{D}\frac{r^2}{a^2(z)}\right)^{\frac{D(\gamma-1)}{4}-\frac{1}{2}}\label{new_uz}.
\end{eqnarray}
Applying the boundary conditions (\ref{bc_n})-(\ref{bc_ur}), one gets the following relationships:
\begin{eqnarray}
 n_c(0)&=&F^{-\frac{D}{2}}\\
 C&=&\frac{Da(0)^2}{R^2}\left(F(0.01-1)\right)^{\frac{2}{D}}\label{new_C}\\
 u_c&=&\frac{a(0)}{a'(0)}\left(F+\frac{C}{Da(0)^2}\right)^{\frac{1}{2}-\frac{D(\gamma-1)}{4}}\label{new_uc}
\end{eqnarray}
It follows that four parameters can be freely chosen between $D$, $F$, $a(0)$, $a'(0)$, and $u_c$.
In order to simplify the study and to be consistent with the previous models, we fix $a'(0)=0.2$, $F=1$, and we vary $D$ and $u_c$ (solving Eqs.(\ref{new_C})-(\ref{new_uc}) for $a(0)$ and $C$).
Note that the profiles of $n$, $u_r$, and $u_z$ at the injection boundary $z=0$ do not depend on $a'(0)$.\\
Figure \ref{fig:profile_new_n} shows the density profile at $z=0$ for $D=-7$ (green), -3 (black), 1 (red), 5 (blue), as a function of $r$.
Clearly, the value of D controls the width of the injected density, that can vary from Gaussian-like to a profile closer to a step-function.
Also, similar to the KT solution, the density tends to zero asymptotically when $D<0$, for $r\rightarrow\infty$.
In this case, however, the value of D determines the asymptotic scaling. 
On the other hand, for $D>0$, one has imaginary values that must be artificially set equal to zero.
Figures \ref{fig:profile_new_ur} and \ref{fig:profile_new_uz} show the 
profiles of $u_r$ and $u_z$, respectively, for $D=-7$ (green), -3 (black), 1 (red), 5 (blue), obtained with $u_c=2$.
Note that whilst the profiles of $n$ and $u_r$ do not change
by varying $u_c$, the axial velocity $u_z$ is a linear function of $u_c$, i.e., the profiles in Figure \ref{fig:profile_new_uz} are linearly rescaled by changing the value of $u_c$.
In Figure \ref{fig:profile_new_D_5_F_1} we present a representative example of the two-dimensional density profile, in logarithmic scale, for $u_c=2$, 10, 20, 100, and $D=-5$.
In order to show the versatility of this class of solutions, we show more examples of the two-dimensional density profile in Figure \ref{fig:profile_new_D_vary}, for $u_c=20$ and varying $D=-7$, -3,
1, 5. Clearly, the cases with $D<0$ do not present a propagation cone.\\
Figures \ref{fig:error_new_z} and \ref{fig:error_new_r} present the errors $\varepsilon_z$ and $\varepsilon_r$, as a function of $u_c$, for $D=-7$ (blue), -3 (green), 5 (red).
As anticipated, the errors are comparable to the ones for the PK and AF solutions, and are monotonically decreasing with increasing $u_c$.

\section{Extension to the case $u_\theta\neq0$}
\begin{figure}
\noindent\includegraphics[width=20pc]{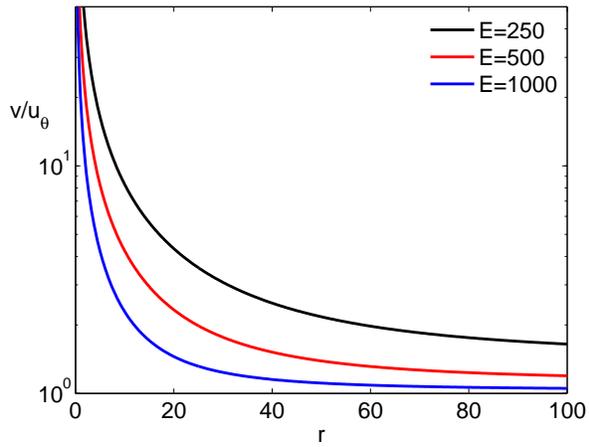}
 \caption{Ratio $v/u_\theta$ as a function of $r$ at $z=0$, for $E=250$ (black), 500(red), and 1000 (blue).}\label{fig:ratio_v_utheta}
\end{figure}

\begin{figure}
\noindent\includegraphics[width=40pc]{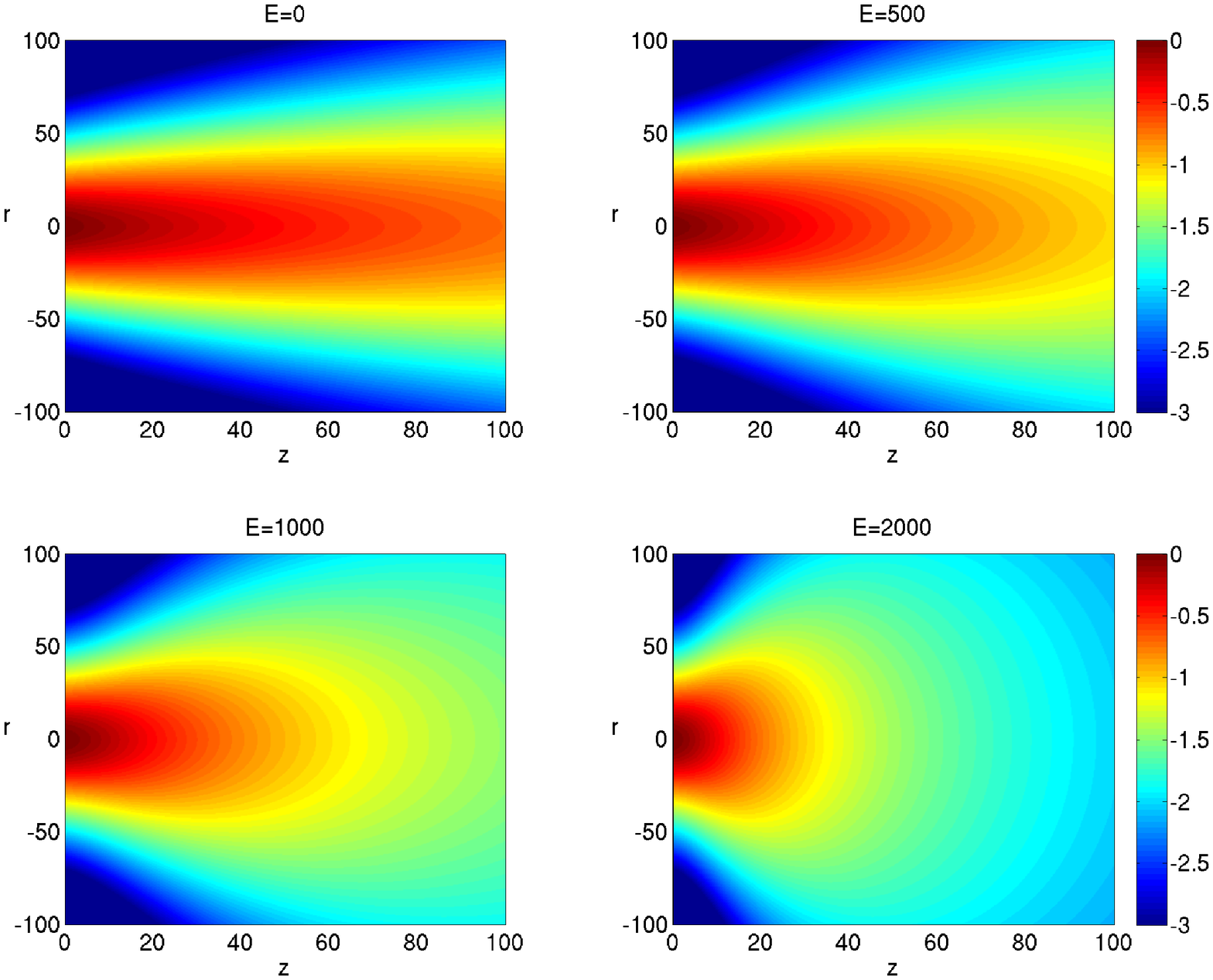}
 \caption{Generalized solution. Two-dimensional profiles of the density in $(z,r)$, for $u_\theta\neq0$, $D=-10$, $F=1$, and $E=0$, 500, 1000, 2000, in logarithmic scale. }\label{fig:profile_theta_1}
\end{figure}

\begin{figure}
\noindent\includegraphics[width=40pc]{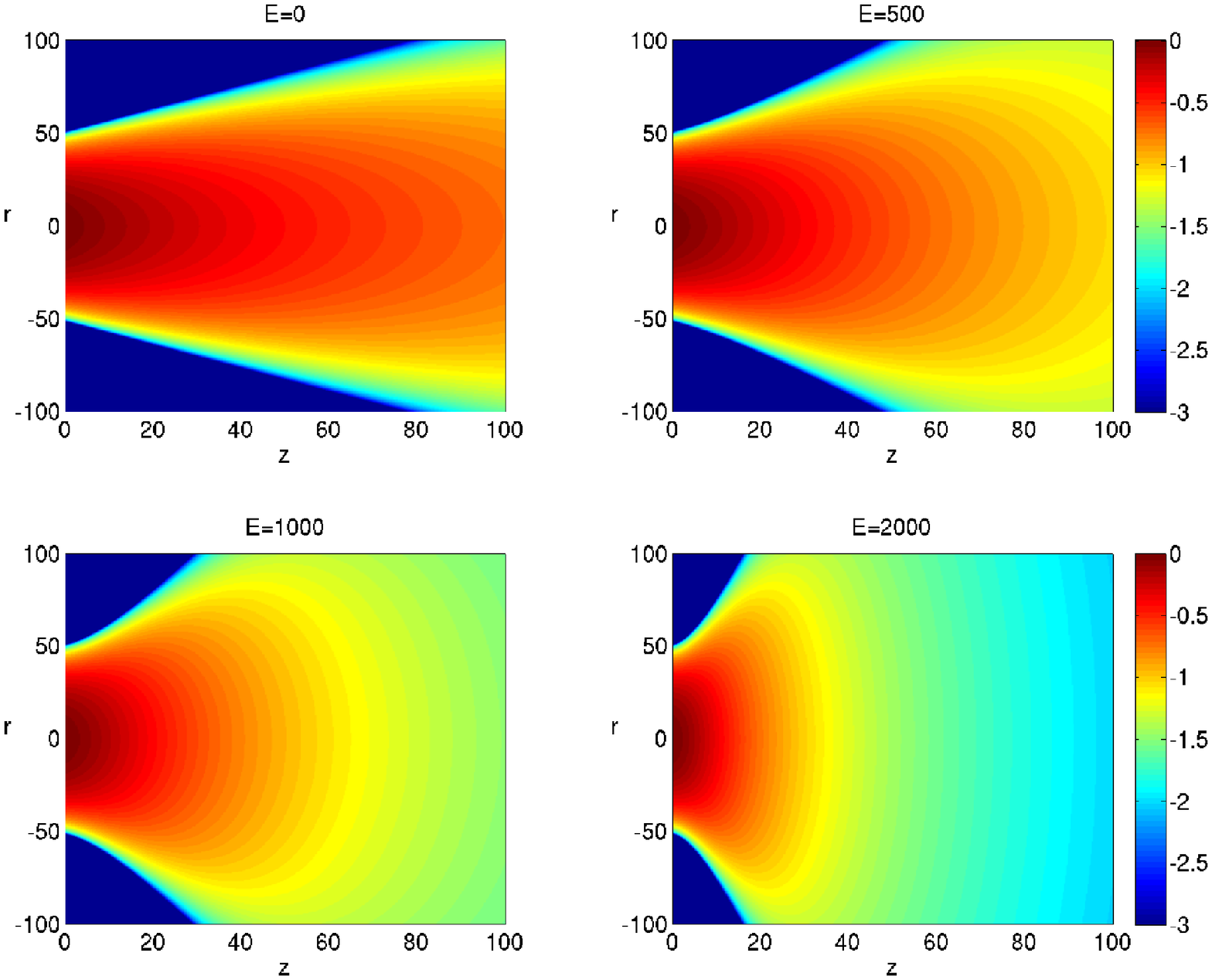}
 \caption{Generalized solution. Two-dimensional profiles of the density in $(z,r)$, for $u_\theta\neq0$, $D=3$, $F=1$, and $E=0$, 500, 1000, 2000, in logarithmic scale. }\label{fig:profile_theta_2}
\end{figure}

By still considering a constant $u_c$, Eq. (\ref{eq_a}) is modified as:
\begin{equation}
 \frac{u_c^2aa''}{n_c^{\gamma-1}}-\frac{E^2}{a^2n_c^{\gamma-1}}=\gamma C.
\end{equation}
Clearly, from Eq.(\ref{eq_utheta}), the value of $E$ determines the magnitude of the azimuthal velocity. We plot in Figure \ref{fig:ratio_v_utheta} the ratio between the total velocity
$v=(u_r^2+u_z^2+u_\theta^2)^{1/2}$ and the azimuthal velocity $u_\theta$, at the injection boundary $z=0$, as a function of $r$, for different values of $E$. Note that $u_\theta$
increases linearly with $r$, and $v/u_\theta$ does not depend on $D$.
The inclusion of a non-null azimuthal velocity at injection causes a distortion of the density profile. Figure \ref{fig:profile_theta_1} shows the two-dimensional density profile in $(z,r)$,
for $E=0$, 500, 1000, 2000, $D=-10$, and $F=1$. The notable effect of the azimuthal velocity is that the density decreases much more sharply along the axial direction, when $u_\theta$ is sufficiently large.
This effect is even more evident when $D>0$, i.e., when the plume is bounded by a propagation cone.
Figure \ref{fig:profile_theta_2} shows the two-dimensional density profile in $(z,r)$ for $D=3$ and $F=1$ (i.e. for the PK solution). Clearly, the cone of propagation is distorted and 
the isocontour $n=0$ becomes more and more curved, with increasing $E$.

\section{Conclusions}
We have analyzed in detail the self-similar solutions presented in \cite{parks79,korsun97,ashkenazy01}, that represent an approximation solution 
for the fluid modeling
of the steady-state, axisymmetric expansion of an electrostatic, quasi-neutral, collisionless plasma plume.
It is important to remind that the semi-analytical solution can be computed in a negligible fraction of the time needed to obtain a full numerical solution,
and as such it is often preferable, as long as we have a measure for the errors. Therefore, in order to judge whether such approximate solutions can reasonably be used in real applications, we have numerically solved the underlying model and we have calculated errors between the approximate
and the numerical solutions. A common characteristic of all the solutions is that the errors decrease with increasing axial injection velocity $u_c$.
For the PK \cite{parks79} and AF \cite{ashkenazy01} solutions, the errors are smaller than $1\%$ for $u_c\gtrsim20$. The KT solution yields larger errors. Each solution is characterized by a given profile for the density
and velocity at injection, and the user has little freedom for adjusting the profiles for more general situations.
For this reason, we have introduced a new class of solutions, that includes the PK and the KT solutions as special cases. We have shown that this new family of solutions can describe fairly general injection profiles,
 yielding errors comparable to the PK and AF models.
Finally, we have shown that a fairly simple extension to the case of a non-null azimuthal velocity $u_\theta$ is possible.
Interestingly, the effect of the azimuthal velocity is to distort the geometry of the plume in such a way that the density decreases more sharply in the vicinity of the 
injection boundary. In real applications, such a scenario might be more favorable for plasma contactors, where it is preferable to have the plume as close as possible to its source, in order to maximize 
the efficiency of the charge neutralization process.

%
%

\ack
We thank G.L. Delzanno, J.E. Borovsky, and M.F. Thomsen for useful discussions.
This work was partially funded by the Laboratory Directed Research
and Development program (LDRD), under the auspices of the
National Nuclear Security Administration of the U.S. Department of
Energy by Los Alamos National Laboratory, operated by Los Alamos
National Security LLC under contract DE-AC52-06NA25396.

\section*{Appendix: Derivation of Equations (\ref{z_cont})-(\ref{r_momr})}
Substituting Eqs. (\ref{eq_n})-(\ref{eq_ur}) in the continuity equation (\ref{continuity}) yields:
\begin{equation}
 2n_cu_ca'+a(u_cn_c'+n_cu_c')=0,
\end{equation}
which reduces to 
\begin{equation}
 (a^2n_cu_c)'=0
\end{equation}
from which Eq. (\ref{z_cont}) follows. For what concerns the momentum equation in the $r$ direction (\ref{mom_r}) one gets:
\begin{equation}
 \frac{\gamma(n_cn_t)^\gamma n_t'}{n_cn_t^2}+\eta au_cu_t^2(a'u_c'+u_ca'')=0,
\end{equation}
that can be separated as
\begin{equation}
 \frac{u_ca}{n_c^{\gamma-1}}\left(u_ca'\right)'=-\frac{\gamma n_t^{\gamma-2}n_t'}{\eta u_t^2}=\gamma C,
\end{equation}
from which Eqs. (\ref{z_momr}) and (\ref{r_momr}) follow.\\
Substituting Eqs. (\ref{eq_n})-(\ref{eq_ur}) in the momentum equation in the $z$ direction (\ref{mom_z}) one obtains:
\begin{equation}
 \gamma(n_cn_t)^{\gamma-2}\left(n_tn_c'-\frac{a'}{a}\eta n_cn_t'\right)+u_cu_t^2u_c'=0,\label{app_z}
\end{equation}
which is unfortunately not separable, and this is the source of the over determination of the system of Eqs. (\ref{z_cont})-(\ref{r_momr}).
In order to make Eq.(\ref{app_z}) separable, one has to make the assumption
\begin{equation}
 \gamma n_c^{\gamma-2}n_c'+u_cu_c'=0
\end{equation}
that can be integrated to obtain Equation (\ref{z_momz}). Finally, using this assumption and taking the $\eta$ derivative of Eq.(\ref{app_z}), the result separates as
\begin{equation}
 \frac{n_c'a}{n_ca'}=-\frac{2n_t'u_t}{2n_tu_t'-(\gamma-1)n_t'u_t}=D
\end{equation}
that ultimately yields, by integration, Eqs. (\ref{z_momz2}) and (\ref{r_momz}).

\newpage
\section*{References}
%
%
%
%
%
%
%
%


\newpage
%
%
%

\clearpage

\end{document}